\documentclass[preprint,11pt]{elsarticle}

\usepackage[T1]{fontenc}
\usepackage[utf8]{inputenc}
\usepackage{lmodern}
\usepackage[a4paper,margin=1in]{geometry}
\usepackage{amsmath,amssymb,amsfonts,amsthm,mathtools,bm}
\usepackage{physics}
\usepackage{graphicx}
\usepackage{xcolor}
\usepackage{booktabs}
\usepackage{enumitem}
\usepackage{hyperref}
\usepackage[nameinlink,capitalise,noabbrev]{cleveref}
\usepackage{microtype}
\usepackage{float}
\usepackage{subcaption}
\usepackage{array}
\usepackage{multirow}
\usepackage{tikz}
\usepackage{url}
\usepackage{doi}

\hypersetup{
  colorlinks=true,
  linkcolor=blue!50!black,
  citecolor=blue!50!black,
  urlcolor=blue!50!black,
  pdftitle={A Wigner-based volumetric transport framework for paraxial waves in random media},
  pdfauthor={Author 1, Author 2}
}

\theoremstyle{definition}

\newcommand{\R}{\mathbb{R}}
\newcommand{\E}{\mathbb{E}}
\newcommand{\I}{\mathcal{I}}
\newcommand{\Jm}{\mathcal{J}}
\newcommand{\W}{\overline{W}}
\newcommand{\Wpsi}{W_\psi}
\newcommand{\Qkin}{\mathcal{Q}_{\mathrm{kin}}}
\newcommand{\Deff}{D_{\mathrm{eff}}}
\newcommand{\atm}{\mathrm{atm}}
\newcommand{\dif}{\mathop{}\!\mathrm{d}}
\newcommand{\trans}{\mathsf{T}}
\renewcommand{\Tr}{\operatorname{Tr}}

\newcommand{\eps}{\varepsilon}
\newcommand{\Cov}{Cov}
\newcommand{\rr}{\mathbf{r}}
\newcommand{\pp}{\mathbf{p}}
\newcommand{\yy}{\mathbf{y}}

\journal{Wave Motion}

\begin{document}

\begin{frontmatter}

\title{A Wigner-based volumetric transport framework for paraxial waves in random media}

\author[aff1]{Arnaud COATANHAY}
\ead{arnaud.coatanhay@ensta.fr}

\author[aff2]{Thomas BONNAFONT}
\ead{thomas.bonnafont.enac@gmail.com}

\author[aff1]{Angélique DREMEAU}
\ead{angelique.dremeau@ensta.fr}

\address[aff1]{Lab-STICC, UMR CNRS 6285, ENSTA, Institut Polytechnique de Paris, \newline
2 rue Fran\c{c}ois Verny, 29806 Brest Cedex 9, France}
\address[aff2]{DGA, 60 Bd. du général Martial Valin, Paris 75509, France}

\begin{abstract}
We develop a Wigner-based phase-space framework for mean paraxial wave propagation in random media. Starting from the random parabolic wave equation, we derive the exact evolution of the realization-dependent Wigner distribution and identify the ensemble-averaged Wigner function as the natural second-order state variable. The averaged equation contains a closure defect, given by a mixed field--medium correlation, which prevents a closed transport equation from being obtained without additional assumptions. We therefore organize the modelling as a hierarchy from the random wave equation to an exact Wigner formulation, then to a nonlocal kinetic closure, and finally to a local Fokker--Planck reduction in the small-angle regime. For the minimal homogeneous isotropic Fokker--Planck model, we derive closed evolution laws for the quadratic moments, exhibit the cubic-in-distance contribution to beam spreading, and obtain explicit Gaussian and Gauss--Schell propagation formulas. These analytical results are used to validate a phase-space splitting solver in one-dimensional transverse benchmarks. Comparisons with nonlocal kinetic models show that the diffusive approximation is accurate for narrow momentum-transfer kernels and loses validity in a controlled way as finite-jump effects become significant. Finally, we introduce a first atmospheric specialization based on a regularized turbulence spectrum, yielding an effective diffusion coefficient expressed in terms of standard atmospheric parameters.
\end{abstract}


\end{frontmatter}

\section*{Introduction}

Wave propagation in random media, such as turbulent ones, is a topic of major interest in radio-wave propagation~\cite{tsang1987radiative,ishimaru1991wave,lin2006influence,kintner2009gnss,darchy2024theoretical}, acoustic propagation~\cite{wilson2001statistical,blanc2002propagation,flatte2005wave} or geoscience~\cite{korn1993seismic,ryzhik1996transport,shapiro1999elastic}. Indeed, the stochastic effects associated shall be modeled to improve the imaging techniques~\cite{garnier2008coherent,borcea2021imaging,morel2024sar} and or the communications~\cite{sinha2023capacity} by introducing an accurate {\it a priori}. In particular, here we consider the refractive index to be a random field in the domain. This models the case of turbulent media such as the atmosphere, troposphere~\cite{darchy2024theoretical} and ionosphere~\cite{vasylyev2022modeling,morel2025ionospheric}, for radio-communications. In this context, the literature is quite dense, from asymptotic methods~\cite{rytov1989principles,morel2025ionospheric}, or Monte-Carlo based ones~\cite{grimault1998multiple,carrano2011multiple,fabbro2013scintillation,darchy2024theoretical}, to stochastic differential equations theory~\cite{wilson2001statistical,garnier2014scintillation}.

Indeed, in the case of a thin layer of random media compared to the propagation distance, one can use the Rytov's approximation~\cite{tatarskii1971effects,rytov1989principles}. The latter has been shown to works well for ionospheric propagation under weak turbulence regime and either plane wave or spherical wave~\cite{morel2025ionospheric} illumination assumptions. In addition, it has the advantage to lead to analytical formulas leading to real time calculations. Nonetheless, this requires to assume the media to be thin and modeled as a phase screen, and to be restricted to the weak turbulence regime. On the other hand, the multiple phase screen method has been developed to overcome these problems. Indeed, the latter is based on the parabolic wave equation (PWE) model~\cite{levy2000parabolic}, which is widely used for long-range propagation, and efficiently solved numerically using the split-step Fourier method~\cite{grimault1998multiple,levy2000parabolic}. The idea is then to construct numerous sample of the random atmosphere to calculate the associated values of the field that yields estimate of the statistical moments~\cite{darchy2024theoretical}. This approach is not limited to the weak scattering regime but rely on a Monte-Carlo approach that is known to have a slow convergence. 

Another strand of the literature focuses on  rewriting the PWE as a stochastic differential equation~\cite{wilson2001statistical,fannjiang2004scaling,garnier2014scintillation,garnier2015white,borcea2016derivation,garnier2023wave,borcea2023paraxial}, thus modeling the random refractive index as a multiplicative noise. In particular, the white-noise model has been widely developed~\cite{garnier2015white,borcea2016derivation}. Indeed, we can then rewrite the PWE as an Itô--Schrödinger equation. One can then obtain a set of equations for the different statistical moments. For the mean, this directly lead to a damping due to the stochastic nature of the refractive index. For the second and fourth order moment, the Wigner transform is usually used to derive solutions~\cite{garnier2014scintillation,garnier2015white,garnier2023wave} for Gaussian source and usual turbulence models. These developments are equivalent to the Markov-PWE moment equations, that are usually solved numerically. This has also helped to introduce new imaging techniques. Nonetheless, in all these cases the random structure, the closure of the stochastic equation, is chosen beforehand. While it has some advantages since direct stochastic equation theory can be used, it can lack physical interpretability. As an example in turbulent plasma modeling the closure defect problematic is widely studied~\cite{weber2018paraxial}.

In this article, as in~\cite{bal2003self,bal2004self,blakaj2026reflection}, we make the choice of studying the Wigner distribution of the second-order moment of the field. Indeed, the phase-space representation it allows is particularly suited here, since it can be seen as an in-between among full-ray solver and ray tracing. It provides an accurate representation of the mean intensity, angular spreading and beam width while being lighter than solving the wave equation for numerous realization of the media. In particular, this approach has proven to be fruitful for the description of rough surface scattering~\cite{cecconi2019diffusive,blakaj2026reflection} where in addition a dynamical energy analysis (DEA) solution is proposed. This shows that this approach could be used in electromagnetic propagation. Nonetheless, in the case of turbulence modeling one shall derive a volume version of the DEA.

This article thus proposes to derive the exact random PWE for the Wigner distribution and highlight the closure defect we have when considering its ensemble average. Following~\cite{kuehn2016moment,weber2018paraxial}, we then propose some different strategies to overcome the closure defect by using physical assumptions instead of a direct noise model. In particular, we introduce a non-local kinetic model that is then reduced to a local Fokker-Planck equation using the Kramers--Moyal expansion. This strategy allows to highlight where the equation is exact and where assumptions are made and under which regime they are valid. From the Fokker-Planck equation, we also obtain a minimal model, for isotropic random refractive index, that allows us to prove a cubic-in-distance contribution to the beam-width as well as a Gaussian propagation theory.  In addition, a volumetric DEA numerical solutions is proposed based on a split-step method. The latter is validated by a comparison with the calculated fundamental solutions, i.e. for Gaussian and Gauss-Schell initial fields. Finally, a numerical experiment is proposed for a realistic turbulent atmosphere case.

\section{Theoretical model}
\label{sec:theo-mod}
\subsection{Exact framework, state variable, and modeling target}
\label{sec:exact-framework}

We consider the propagation of a quasi-monochromatic scalar optical field through a weakly heterogeneous random medium, predominantly along a distinguished longitudinal axis $z$. The physically relevant dynamics takes place in the transverse variable $\rr \in\R^d$,
with $d=2$ in the standard paraxial beam-propagation setting. For a fixed realization of the medium, the complex envelope $\psi(z,\rr)$ is assumed to satisfy the random narrow angle paraxial wave equation~\cite{garnier2015white}
\begin{equation}
2ik_0\,\partial_z \psi(z,\rr)+\Delta_{\rr}\psi(z,\rr)+2k_0^2\eta(z,\rr)\psi(z,\rr)=0,
\label{eq:paraxial-random}
\end{equation}
where $k_0>0$ is the reference wavenumber, $\Delta_{\rr}$ denotes the transverse Laplacian, and $\eta(z,\rr) \in \mathcal{L}^2(\R)$ models the refractive-index perturbation relative to a homogeneous background, i.e. $n(z,\rr) = 1+\eta(z,\rr)$. We then write
\begin{equation}
\eta(z,\rr)=\bar{\eta}(z,\rr)+\delta\eta(z,\rr),
\label{eq:eta-split}
\end{equation}
where $\bar{\eta}$ is a deterministic slowly varying component and $\delta\eta$ is a centered fluctuation such that $ \E[\delta\eta(z,\rr)]=0$.
Contrary to~\cite{garnier2014scintillation,garnier2015white}, here no closure assumption are introduced, i.e. we do not make any approximation of the noise shape.

The random field $\delta\eta$ is only assumed to possess well-defined second-order statistics. We thus only require the covariance function
\begin{equation}
R(\zeta,\mathbf{\rho};z,\rr)
=
\E\!\left[
\delta\eta(z+\zeta,\rr+\mathbf{\rho})\,\delta\eta(z,\rr)
\right]
\label{eq:general-covariance}
\end{equation}
to be well defined. No stationarity, isotropy, or specific spectral model is imposed at this point, which is essential to the logic of the paper. Indeed the purpose is not to specialize immediately to a turbulence model, but to identify the exact second-order state variable from which the hierarchy of closures will be constructed~\cite{kuehn2016moment,weber2018paraxial}.

In practice, systems measures the illumination, which correspond to $|\psi|^2$, i.e. the power. We are thus only interested in statistical moments of order more than $2$. Over the years, the Wigner transform has proven to be useful to calculate second order statistics in electromagnetics~\cite{bal2003self,bal2004self}. Therefore,  for each realization of the medium, we define here the transverse Wigner distribution as
\begin{equation}
\Wpsi(z,\rr,\pp)
=
\left(\frac{k_0}{2\pi}\right)^d
\int_{\R^d}
\psi\!\left(z,\rr+\frac{\yy}{2}\right)
\psi^*\!\left(z,\rr-\frac{\yy}{2}\right)
e^{-ik_0 \pp \cdot \yy}\,\dif \yy,
\label{eq:wigner-realization}
\end{equation}
where $\pp\in\R^d$ is the transverse phase-space variable conjugate to $\yy$, and $^*$ denotes the complex conjugate. This quantity provides a phase-space representation of second-order wave information for a single realization. As usual, it is not in general pointwise positive and may retain oscillatory interference structure since it is only a quasi-probability distribution. In this work, the studied state variable is not \eqref{eq:wigner-realization} itself, but its ensemble average
\begin{equation}
\W(z,\rr,\pp):=\E[\Wpsi(z,\rr,\pp)].
\label{eq:averaged-wigner}
\end{equation}
The aim of the paper is to describe mean second-order propagation in random media, not the instantaneous complex field associated with a single realization. In that sense, $\W$ plays the role of an effective phase-space density for ensemble-averaged transport. From the latter, two main macroscopic observables can be defined
\begin{equation}
\I(z,\rr):=\int_{\R^d}\W(z,\rr,\pp)\,\dif \pp,
\label{eq:spatial-marginal}
\end{equation}
and
\begin{equation}
\Jm(z,\pp):=\int_{\R^d}\W(z,\rr,\pp)\,\dif \rr.
\label{eq:angular-marginal}
\end{equation}
They correspond to the mean transverse irradiance and the mean angular distribution, respectively. These marginals provide the direct link between the phase-space formulation and observable beam quantities. In particular, spatial second moments of $\I$ will be used to characterize beam spreading, while angular second moments of $\Jm$ will quantify directional broadening. 

The modeling target may now be stated precisely. Starting from the exact random paraxial model \eqref{eq:paraxial-random}, we seek an effective closed evolution equation for the averaged Wigner distribution $\W(z,r,p)$. Since no closure is assumed at first, the construction rely on three steps~\cite{kuehn2016moment}:  first we derive the exact Wigner evolution for a single realization, then we average and identifies the corresponding closure defect, third we replace that defect by an effective nonlocal kinetic operator. In particular, we choose to only study these second-order quantity in this paper and do not attempt to characterize the variance of the Wigner distribution, i.e. the scintillations.


Finally, the present approach is conceptually related to Wigner-based transport methods developed for rough-surface scattering~\cite{blakaj2026reflection} and other boundary-driven problems. The essential difference is that the disorder considered here is volumetric rather than localized on interfaces. As a consequence, the phase-space evolution is continuous in $z$ rather than formulated as a stationary transfer problem between boundaries. The present work should therefore not be viewed as a direct extension of DEA~\cite{lyon2014theory,cecconi2019diffusive,blakaj2026reflection} in the strict boundary-operator sense, but rather as a volumetric phase-space transport framework guided by the same principle.

\subsection{Exact Wigner equation and closure defect}
\label{sec:exact-wigner}
Starting from the random paraxial wave equation \eqref{eq:paraxial-random}, we obtain the following equation for the Wigner distribution
\begin{equation}
\partial_z \Wpsi(z,\rr,\pp)
+
\pp \cdot \nabla_{\rr} \Wpsi(z,\rr,\pp)
=
\Theta_{\eta}[\Wpsi](z,\rr,\pp),
\label{eq:exact-wigner}
\end{equation}
where the medium operator is
\begin{equation}
\Theta_{\eta}[\Wpsi](z,\rr,\pp)
=
i k_0
\left(\frac{k_0}{2\pi}\right)^d
\int_{\R^d}
\Bigl[
\eta\!\left(z,\rr+\frac{\yy}{2}\right)
-
\eta\!\left(z,\rr-\frac{\yy}{2}\right)
\Bigr]
\Gamma_\psi(z,\rr,\yy)\,
e^{-ik_0 \pp \cdot \yy}\,\dif \yy,
\label{eq:moyal-operator-real-space}
\end{equation}
where $\Gamma_\psi(z,\rr,\yy)=\psi\!\left(z,\rr+\frac{\yy}{2}\right)
\psi^*\!\left(z,\rr-\frac{\yy}{2}\right)$. The calculation details are given in~\ref{app:wigner-derivation}. Note that equation \eqref{eq:exact-wigner} is exact. Its structure is already informative: the left-hand side is the free transport operator in transverse phase space, whereas all medium effects are gathered in the nonlocal operator $\Theta_\eta$. We can rewrite the latter in the usual Fourier variables as
\begin{equation}
\Theta_{\eta}[\Wpsi](z,\rr,\pp)
=
\frac{i k_0}{(2\pi)^d}
\int_{\R^d}
\widehat{\eta}(z,\mathbf{q})\,e^{i\mathbf{q}\cdot \rr}
\left[
\Wpsi\!\left(z,\rr,\pp-\frac{\mathbf{q}}{2k_0}\right)
-
\Wpsi\!\left(z,\rr,\pp+\frac{\mathbf{q}}{2k_0}\right)
\right]
\dif \mathbf{q}.
\label{eq:moyal-operator-fourier}
\end{equation}
This makes the momentum-transfer structure explicit: the medium acts through opposite shifts in the variable $\pp$, with typical jump size of order $|\mathbf{q}|/k_0$. 

Then, using the decomposition $\eta=\bar{\eta}+\delta\eta$, equation \eqref{eq:exact-wigner} becomes
\begin{equation}
\partial_z \Wpsi (z,\rr,\pp) + \pp\cdot \nabla_{\rr} \Wpsi (z,\rr,\pp)
=
\Theta_{\bar{\eta}}[\Wpsi] (z,\rr,\pp)
+
\Theta_{\delta\eta}[\Wpsi] (z,\rr,\pp).
\label{eq:exact-wigner-split}
\end{equation}
The term $\Theta_{\bar{\eta}}$ represents the contribution of the resolved deterministic part of the medium, while $\Theta_{\delta\eta}$ is the random scattering mechanism. 
Taking expectations in \eqref{eq:exact-wigner-split}, we obtain
\begin{equation}
\partial_z \W(z,\rr,\pp)
+
\pp \cdot \nabla_{\rr} \W(z,\rr,\pp)
=
\Theta_{\bar{\eta}}[\W](z,\rr,\pp)
+
\mathcal{C}(z,\rr,\pp),
\label{eq:averaged-exact-wigner}
\end{equation}
where the exact closure defect is
\begin{equation}
\mathcal{C}(z,\rr,\pp)
:=
\E\!\left[\Theta_{\delta\eta}[\Wpsi](z,\rr,\pp)\right].
\label{eq:closure-defect}
\end{equation}
Equation \eqref{eq:averaged-exact-wigner} is still exact, but it is not closed, since $\mathcal{C}$ cannot in general be expressed as an operator acting on $\W$ alone, which is the main difference from~\cite{garnier2014scintillation} where a white-noise model of the random refractive index is assumed closing the system. The structure of this defect is made explicit by writing
\begin{equation}
\mathcal{C}(z,\rr,\pp)
=
i k_0
\left(\frac{k_0}{2\pi}\right)^d
\int_{\R^d}
\E\!\left(
\Bigl[
\delta\eta\!\left(z,\rr+\frac{\yy}{2}\right)
-
\delta\eta\!\left(z,\rr-\frac{\yy}{2}\right)
\Bigr]
\Gamma_\psi(z,\rr,\yy)
\right)
e^{-ik_0 \pp\cdot \yy}\,\dif \yy,
\label{eq:closure-defect-real-space}
\end{equation}
or, equivalently, in Fourier form,
\begin{equation}
\mathcal{C}(z,\rr,\pp)
=
\frac{i k_0}{(2\pi)^d}
\int_{\R^d}
\E\!\left[
\widehat{\delta\eta}(z,\mathbf{q})\,e^{i\mathbf{q}\cdot \rr}
\left(
\Wpsi\!\left(z,\rr,\pp-\frac{\mathbf{q}}{2k_0}\right)
-
\Wpsi\!\left(z,\rr,\pp+\frac{\mathbf{q}}{2k_0}\right)
\right)
\right]
\dif \mathbf{q}.
\label{eq:closure-defect-fourier}
\end{equation}
This is the central modeling obstacle of the paper: the exact averaged dynamics involves a mixed medium--field correlation and therefore depends on more information than the averaged Wigner function alone, i.e. on higher order moments~\cite{kuehn2016moment}. In other words, ensemble averaging does not automatically produce a closed transport equation for $\W$.

It is also useful to reinterpret the same construction from the viewpoint of a volumetric transfer theory. For a fixed realization of the medium, the paraxial field may be written formally in terms of a Green propagator~\cite{palmer1979path}
\begin{equation}
\psi(z,\rr)
=
\int G(z,z_0;\rr,\mathbf{\rho})\,\psi(z_0,\mathbf{\rho})\,\dif \mathbf{\rho}.
\label{eq:green-field-representation}
\end{equation}
In particular, for the deterministic case, the propagator is known and correspond to a multiplication of two exponential: one accounting for the diffusion, the other that accounts for the refractive index. At the field level this representation is exact, but, in our case, the relevant transport object is not the Green function itself. Indeed, it is rather the second-order propagator associated with the field correlation function,
\begin{equation}
\Gamma(z;\rr_1,\rr_2)
=
\mathbb{E}\!\left[\psi(z,\rr_1)\psi^*(z,\rr_2)\right],
\label{eq:gamma-second-order}
\end{equation}
namely
\begin{equation}
K_2(z,z_0;\rr_1,\rr_2;\mathbf{\rho}_1,\mathbf{\rho}_2)
:=
\mathbb{E}\!\left[
G(z,z_0;\rr_1,\mathbf{\rho}_1)\,
G^*(z,z_0;\rr_2,\mathbf{\rho}_2)
\right].
\label{eq:K2-def}
\end{equation}
Passing from $\Gamma$ to its Wigner transform yields an induced transfer operator in phase space,
\begin{equation}
W(z)=\mathcal{T}_{z,z_0}W(z_0),
\label{eq:phase-space-transfer}
\end{equation}
which may be viewed as the volumetric analogue of the transfer operators used in DEA for boundary-to-boundary propagation~\cite{blakaj2026reflection}. The difference is that the present setting is continuous in the propagation coordinate $z$. Instead of a stationary transfer between reflections, one obtains a propagation semi-group acting along the medium. For sufficiently small propagation increments $\Delta z$, one expects
\begin{equation}
\mathcal{T}_{z+\Delta z,z}
=
I+\Delta z\,\mathcal{L}+o(\Delta z),
\label{eq:generator-expansion}
\end{equation}
with effective generator
\begin{equation}
\mathcal{L}
=
-\pp\cdot \nabla_{\rr} + \mathcal{Q}.
\label{eq:generator-form}
\end{equation}
which is the analogue of the split-step operator of the deterministic case. Indeed, for very small propagation steps, the exponential can be approximated by its argument~\cite{braidotti2018path}. 

This analysis gives a clear justification to the closure hierarchy used in the remainder of the article. Indeed, the free transport term describes the deterministic part of phase-space propagation, while the operator $\mathcal{Q}$ encodes the cumulative action of the random medium. Thus, keeping $\mathcal{Q}$ in nonlocal form leads to the kinetic closure, which is introduced in details in Section~\ref{sec:kinetic-closure}. Then, assuming a second-order small-jump expansion, we derive the local Fokker--Planck reduction, developed in Section~\ref{sec:fp-model}. The present framework may thus be understood as a continuous and volumetric Wigner-based DEA. 

\subsection{Asymptotic assumptions and nonlocal kinetic closure}
\label{sec:kinetic-closure}
Now that the closure defect~\eqref{eq:closure-defect-real-space} has been highlighted, i.e. $\mathcal{C}(z,r,p)=\E\!\left[\Theta_{\delta\eta}[\Wpsi](z,r,p)\right]$, we shall find an approximation of the latter as an operator that acts only on $\W$. Following the hierarchical strategy for moment closure~\cite{kuehn2016moment} and the work of Weber {\it et al.}~\cite{weber2018paraxial}, we first introduce a nonlocal kinetic closure in the transverse momentum variable.

The closure construction relies on a standard asymptotic picture. We remain in the paraxial forward regime, assume that the random fluctuation field is weak at the local level but cumulative over propagation distances of interest, and require sufficient longitudinal de-correlation so that the closure defect may be approximated by a Markovian operator in $z$. In addition, the wavelength scale is assumed small compared with the relevant transverse correlation scales of the medium, so that the momentum-shift structure already visible in \eqref{eq:moyal-operator-fourier} may be replaced, after averaging, by an effective jump operator in $p$. Finally, the whole construction is restricted to second-order quantities derived from the averaged Wigner function, that can be seen as neglecting the higher order terms in  Furutsu-Novikov theorem~\cite{bragg2012drift}.

Under these assumptions, we approximate the exact closure defect by a nonlocal scattering operator acting on $\W$. Thus, the corresponding effective phase-space model reads
\begin{equation}
\partial_z \W(z,\rr,\pp)
+
\pp\cdot \nabla_{\rr} \W(z,\rr,\pp)
+
F(z,\rr,\pp)\cdot \nabla_{\pp} \W(z,\rr,\pp)
=
\Qkin[\W](z,\rr,\pp),
\label{eq:kinetic-model}
\end{equation}
where $F$ represents the resolved deterministic refractive drift associated with the mean medium structure, and $\Qkin$ is a nonlocal operator in momentum. The is the paraxial equivalent of the optic equation used in~\cite{maj2026beam}. The most general form retained in this paper is
\begin{equation}
\Qkin[\W](z,\rr,\pp)
=
\int_{\R^d}
\sigma(z,\rr;\pp,\pp')
\left[
\W(z,\rr,\pp')-\W(z,\rr,\pp)
\right]
\,\dif \pp',
\label{eq:collision-operator-sigma}
\end{equation}
with $\sigma(z,\rr;\pp,\pp')\ge 0$ the scattering cross-section, in line with the results derived in the electron-cyclotron wave propagation literature~\cite{weber2015scattering,weber2018paraxial,maj2026beam}. This equation highlights the effect of the turbulent media. There is a redistribution of the density in the direction $\pp$ from other directions $\pp'$, and losses from $\pp$ toward the rest of momentum space. It shall be noted that under a cumulative white-noise assumption for the refractive index, using the Furutsu-Novikov theorem, we retrieve the moment equation of Garnier {\it et al.}~\cite{garnier2014scintillation}, see~\ref{app:equivalence-FN} for the complete proof.

For the asymptotic analysis, it is convenient to rewrite the operator in terms of momentum increments. Introducing the jump variable
$\mathbf{q}=\pp'-\pp$,
equation \eqref{eq:collision-operator-sigma} becomes
\begin{equation}
\Qkin[\W](z,\rr,\pp)
=
\int_{\R^d}
K(z,\rr;\mathbf{q},\pp)
\left[
\W(z,\rr,\pp+\mathbf{q})-\W(z,\rr,\pp)
\right]
\,\dif \mathbf{q},
\label{eq:jump-kernel}
\end{equation}
where $K(z,\rr;\mathbf{q},\pp):=\sigma(z,\rr;\pp,\pp+\mathbf{q})$. It shall be noted that the kernel $K(z,r,q;p)$ should not be interpreted as a microscopic scattering cross section, but more as an effective macroscopic one. 




\subsection{Diffusive reduction and minimal Fokker--Planck model}
\label{sec:fp-model}

We now derive a local diffusive approximation of the nonlocal kinetic model introduced in \cref{sec:kinetic-closure}. Our goal is to identify the asymptotic regime in which the jump operator
\begin{equation}
\Qkin[\W](z,\rr,\pp)
=
\int_{\R^d}
K(z,\rr;\mathbf{q},\pp)
\left[
\W(z,r,\pp+\mathbf{q})-\W(z,\rr,\pp)
\right]
\,\dif \mathbf{q}
\label{eq:jump-operator-again}
\end{equation}
may be approximated by a differential operator in the momentum variable. 

Under the small-angle assumptions, i.e. the kernel is concentrated near $\mathbf{q}=0$ and $\W$ vary slowly and smoothly in the momentum space, the jump operator admits a second-order Kramers--Moyal expansion in the variable $\mathbf{q}$. This writes as
\begin{equation}
\Qkin[\W](z,\rr,\pp)
\approx
b(z,\rr,\pp)\cdot \nabla_{\pp} \W(z,\rr,\pp)
+
\frac12
\sum_{i,j=1}^d
A_{ij}(z,\rr,\pp)\,
\partial_{p_i}\partial_{p_j}\W(z,\rr,\pp),
\label{eq:qm-second-order}
\end{equation}
where
\begin{equation}
b_i(z,\rr,\pp)
:=
\int_{\R^d}
q_i\,K(z,\rr;\mathbf{q},\pp)\,\dif \mathbf{q},
\label{eq:def-bi}
\end{equation}
and
\begin{equation}
A_{ij}(z,\rr,\pp)
:=
\int_{\R^d}
q_iq_j\,K(z,\rr;\mathbf{q},\pp)\,\dif \mathbf{q}.
\label{eq:def-aij}
\end{equation}
The detailed derivation, including the structure of the remainder term, is given in \ref{app:kramers-moyal}. At this stage, the closure is local in $\pp$, but it still allows for considering anisotropy, inhomogeneity, and a nonzero drift in momentum space.

It is often convenient to rewrite the second-order truncation in divergence form as
\begin{equation}
\Qkin[\W]
\approx
-\nabla_{\pp}\cdot\bigl(B(z,\rr,\pp)\,\W(z,\rr,\pp)\bigr)
+
\frac12
\sum_{i,j=1}^d
\partial_{p_i}\partial_{p_j}
\bigl(A_{ij}(z,\rr,\pp)\,\W(z,\rr,\pp)\bigr),
\label{eq:fp-general-div-form}
\end{equation}
for a suitable effective drift field $B$ that depends on the chosen representation. The corresponding complete local drift--diffusion model thus writes
\begin{equation}
\partial_z \W
+
\pp\cdot \nabla_\rr \W
+
F(z,\rr,\pp)\cdot \nabla_\pp \W
=
-\nabla_\pp\cdot\bigl(B(z,\rr,\pp)\,\W\bigr)
+
\frac12
\sum_{i,j=1}^d
\partial_{p_i}\partial_{p_j}
\bigl(A_{ij}(z,\rr,\pp)\,\W\bigr).
\label{eq:fp-general}
\end{equation}
The latter corresponds to the natural local counterpart of the nonlocal kinetic equation~\eqref{eq:kinetic-model}.

An important reference case is when the jump kernel is centered and isotropic, i.e. homogeneous in the transverse variables and isotropic in momentum transfer. In this case, we can further simplify the model~\eqref{eq:fp-general}. More precisely, assume that $K$ is symmetric in $\mathbf{q}$ so that
\begin{equation}
K(z,\rr;\mathbf{q},\pp)=K(z,\rr;-\mathbf{q},\pp),
\label{eq:symmetric-kernel}
\end{equation}
then the first jump moment vanishes,
\[
b(z,\rr,\pp)=0.
\]
Secondly assume that the second jump tensor is isotropic, so that it writes
\begin{equation}
A_{ij}(z,\rr,\pp)=2D(z,\rr,\pp)\,\delta_{ij}.
\label{eq:isotropic-diffusion}
\end{equation}
Then~\eqref{eq:qm-second-order} reduces to
\[
\Qkin[\W]\approx D(z,\rr,\pp)\,\Delta_\pp \W.
\]
If, in addition, the medium is homogeneous in the transverse variables, then $D$ may be taken constant and the diffusive equation becomes
\begin{equation}
\partial_z \W(z,\rr,\pp)
+
\pp\cdot \nabla_\rr \W(z,\rr,\pp)
=
\Deff\,\Delta_\pp \W(z,\rr,\pp),
\label{eq:minimal-fp-model}
\end{equation}
where $\Deff>0$ is the effective angular diffusion coefficient. Equation \eqref{eq:minimal-fp-model} is the minimal model studied in the remainder of the paper. It should be interpreted as the simplest isotropic small-angle closure of the kinetic jump equation, and not as a universal transport law for random media.

The reduction also provides a direct expression for the diffusion coefficient in terms of the second moment of the jump kernel. In the homogeneous isotropic case, we have
\begin{equation}
\Deff
=
\frac{1}{2d}
\int_{\R^d}
|\mathbf{q}|^2 K(\mathbf{q})\,\dif \mathbf{q}.
\label{eq:deff-from-kernel}
\end{equation}
Indeed, isotropy implies
\[
\int_{\R^d} q_iq_jK(\mathbf{q})\,\dif \mathbf{q}
=
\frac{\delta_{ij}}{d}
\int_{\R^d}|\mathbf{q}|^2K(\mathbf{q})\,\dif \mathbf{q},
\]
and identification with~\eqref{eq:isotropic-diffusion} directly yields~\eqref{eq:deff-from-kernel}. When $d=2$, i.e. which is the physical relevant case, this becomes
\begin{equation}
\Deff
=
\frac14
\int_{\R^2}
|\mathbf{q}|^2 K(\mathbf{q})\,\dif \mathbf{q}.
\label{eq:deff-d2}
\end{equation}
In addition, for a radial kernel, i.e. $K(\mathbf{q})=K(|\mathbf{q}|)$, we have
\begin{equation}
\Deff
=
\frac{\pi}{2}
\int_0^\infty
\rho^3 K(\rho)\,\dif \rho.
\label{eq:deff-radial}
\end{equation}
This formula shows that the minimal Fokker--Planck model is not parameterized by an arbitrary constant, but by the normalized second angular moment of the underlying nonlocal redistribution kernel.

From the minimal model~\eqref{eq:minimal-fp-model} we get two immediate important properties. First, when $\Deff\to0$, one recovers the free phase-space transport equation
\[
\partial_z \W + \pp\cdot \nabla_\rr \W = 0,
\qquad
\W(z,\rr,\pp)=\W(0,\rr-z\pp,\pp),
\]
so that the correct deterministic limit is preserved, as expected. Second, for nonnegative integrable initial data, Equation~\eqref{eq:minimal-fp-model} generates a positivity-preserving Markov semigroup, which is consistent with its kinetic interpretation. From the numerical point of view, this limit provides a first validation for the solver: if $\Deff=0$, the numerical scheme must reproduce exact ballistic transport, leave the angular marginal unchanged, broaden the spatial profile without artificial diffusion, and preserve the total mass. 

\section{Some asymptotic results}
\label{sec:asym-res}
\subsection{Quadratic moments and cubic beam-spreading law}
\label{sec:moments}

Throughout this section, we assume that $\W$ is sufficiently regular, nonnegative, and decays fast enough as $|r|+|p|\to\infty$ so that all integrations by parts are justified. We also assume that the total mass
\begin{equation}
M
:=
\int_{\R^d}\int_{\R^d}
\W(z,\rr,\pp)\,\dif \rr\,\dif \pp
\label{eq:mass-def}
\end{equation}
is finite. As already noted in the previous section, $M$ is conserved under \eqref{eq:minimal-fp-model}.

The basic quadratic observables are the spatial second moment
\begin{equation}
R_2(z)
:=
\int_{\R^d}\int_{\R^d}
|\rr|^2\,\W(z,\rr,\pp)\,\dif \rr\,\dif \pp,
\label{eq:R2-def}
\end{equation}
the angular second moment
\begin{equation}
P_2(z)
:=
\int_{\R^d}\int_{\R^d}
|\pp|^2\,\W(z,\rr,\pp)\,\dif \rr\,\dif \pp,
\label{eq:P2-def}
\end{equation}
and the mixed position--momentum moment
\begin{equation}
C(z)
:=
\int_{\R^d}\int_{\R^d}
\rr\cdot \pp\,\W(z,\rr,\pp)\,\dif \rr\,\dif \pp.
\label{eq:C-def}
\end{equation}
These quantities measure, respectively, the mean squared beam width, the mean squared angular spread, and the phase-space correlation between position and direction.

Integrating \eqref{eq:minimal-fp-model} over all phase space immediately gives
\begin{equation}
\frac{\dif M}{\dif z}=0,
\label{eq:mass-conservation}
\end{equation}
since both the transport and diffusion terms vanish by integration by parts. The evolution of the quadratic moments is obtained in the same way. First, multiplying \eqref{eq:minimal-fp-model} by $|\pp|^2$ and integrating over $(\rr,\pp)$, the transport term vanishes and the identity $\Delta_\pp|\pp|^2=2d$ yields
\begin{equation}
\frac{\dif P_2}{\dif z}
=
2d\,\Deff\,M.
\label{eq:P2-evolution}
\end{equation}
Thus the angular variance grows linearly with propagation distance. In the normalized physically relevant case $M=1$ and $d=2$, this reduces to
\begin{equation}
\frac{\dif P_2}{\dif z}=4\,\Deff.
\label{eq:P2-evolution-d2}
\end{equation}
Second, multiplying~\eqref{eq:minimal-fp-model} by $\rr\cdot \pp$ and integrating, the diffusion term vanishes because $\rr\cdot \pp$ is affine in $\pp$, while integration by parts in $\rr$ gives
\begin{equation}
\frac{\dif C}{\dif z}=P_2(z).
\label{eq:C-evolution}
\end{equation}
Finally, multiplying by $|\rr|^2$ and integrating, the diffusion term again vanishes and one finds
\begin{equation}
\frac{\dif R_2}{\dif z}=2C(z).
\label{eq:R2-evolution}
\end{equation}
We thus obtain the closed triangular system
\begin{equation}
\frac{\dif P_2}{\dif z}=2d\,\Deff\,M,
\qquad
\frac{\dif C}{\dif z}=P_2,
\qquad
\frac{\dif R_2}{\dif z}=2C.
\label{eq:moment-system-general}
\end{equation}
It is sometimes useful to rewrite the hierarchy in matrix form,
\begin{equation}
\frac{\dif}{\dif z}
\begin{pmatrix}
R_2\\[0.3em]
C\\[0.3em]
P_2
\end{pmatrix}
=
\begin{pmatrix}
0 & 2 & 0\\
0 & 0 & 1\\
0 & 0 & 0
\end{pmatrix}
\begin{pmatrix}
R_2\\[0.3em]
C\\[0.3em]
P_2
\end{pmatrix}
+
\begin{pmatrix}
0\\[0.3em]
0\\[0.3em]
2d\,\Deff\,M
\end{pmatrix},
\label{eq:moment-matrix-form}
\end{equation}
which emphasizes the one-way cascade structure. Its interpretation is immediate: the medium injects angular variance directly into $P_2$, the angular variance feeds the mixed moment $C$, and the mixed moment in turn drives the spatial spreading $R_2$. 

Finally, let us assume that we have the following initial data
\[
P_2(0)=P_{2,0},
\qquad
C(0)=C_0,
\qquad
R_2(0)=R_{2,0}.
\]
Then, integrating~\eqref{eq:moment-system-general} yields
\begin{equation}
P_2(z)
=
P_{2,0}
+
2d\,\Deff\,M\,z,
\label{eq:P2-explicit}
\end{equation}
\begin{equation}
C(z)
=
C_0
+
P_{2,0}z
+
d\,\Deff\,M\,z^2,
\label{eq:C-explicit}
\end{equation}
and
\begin{equation}
R_2(z)
=
R_{2,0}
+
2C_0 z
+
P_{2,0}z^2
+
\frac{2d}{3}\Deff\,M\,z^3.
\label{eq:R2-explicit}
\end{equation}
Hence the spatial second moment decomposes as
\begin{equation}
R_2(z)
=
R_{2,\mathrm{free}}(z)
+
\frac{2d}{3}\Deff\,M\,z^3,
\label{eq:R2-free-plus-diff}
\end{equation}
where
\begin{equation}
R_{2,\mathrm{free}}(z)
:=
R_{2,0}
+
2C_0 z
+
P_{2,0}z^2
\label{eq:R2-free}
\end{equation}
is the purely ballistic contribution. In the normalized two-dimensional case $M=1$ and $d=2$, this becomes
\begin{equation}
R_2(z)
=
R_{2,\mathrm{free}}(z)
+
\frac{4}{3}\Deff\,z^3.
\label{eq:R2-z3-law-d2}
\end{equation}

The cubic contribution displayed in \eqref{eq:R2-z3-law-d2} is consistent with beam-broadening laws previously obtained in the context of electron-cyclotron wave propagation through turbulent fusion plasmas
\cite{sysoeva2015beam,chellai2021millimeter}. In particular, Chella{\"i} et al.\ derived a more general path-integral expression whose homogeneous limit exhibits the same cubic dependence on propagation distance.

In the present framework, this law follows directly from the closed triangular moment system \eqref{eq:moment-matrix-form}. The random medium acts on the momentum variable and produces a linear growth of the momentum second moment $P_2$. This contribution is successively transferred to the mixed moment $C$ and then to the spatial second moment $R_2$. Consequently, the two integrations involved in the cascade
\[
P_2 \longrightarrow C \longrightarrow R_2
\]
generate a term proportional to $z^3$.

The contribution of the present formulation is therefore not the cubic scaling by itself, but its derivation within a Wigner-based closure hierarchy linking the random wave equation, the nonlocal kinetic model, and its Fokker--Planck reduction.

For centered initial data with vanishing initial position--momentum correlation, $C_0=0$, equation \eqref{eq:R2-explicit} becomes
\begin{equation}
R_2(z)
=
R_{2,0}
+
P_{2,0}z^2
+
\frac{2d}{3}\Deff\,M\,z^3.
\label{eq:R2-centered}
\end{equation}
The first two terms describe the initial and ballistic contributions, whereas the last term isolates the cumulative broadening induced by the random medium. This expression applies naturally to centered Gaussian and Gauss--Schell initial data and will be recovered directly from the explicit solution derived in the next section.



\subsection{Fundamental solution and propagation of Gaussian/Gauss--Schell data}
\label{sec:gaussian-propagation}

We continue with the minimal homogeneous isotropic closure~\eqref{eq:minimal-fp-model} supplemented with the initial condition $\W(0,\rr,\pp)=W_0(\rr,\pp)$. The purpose of this section is twofold: to compute the fundamental solution of \eqref{eq:minimal-fp-model}, and to show that Gaussian phase-space states propagate explicitly under this dynamics.

Equation \eqref{eq:minimal-fp-model} is the Fokker--Planck equation associated with the following linear stochastic system
\begin{equation}
\dif r_z = p_z\,\dif z,
\qquad
\dif p_z = \sqrt{2\Deff}\,\dif B_z,
\label{eq:sde-model}
\end{equation}
where $(B_z)_{z\ge 0}$ is a standard Brownian motion in $\R^d$. This immediately implies that the transition kernel is Gaussian: the momentum variable undergoes diffusion, while the position variable is its time integral. 

Then, for initial data $(\rr_0,\pp_0)$, a direct integration leads to
\begin{equation}
\pp_z=\pp_0+\sqrt{2\Deff}\,B_z,
\qquad
\rr_z=\rr_0+z\pp_0+\sqrt{2\Deff}\int_0^z B_s\,\dif s,
\label{eq:appC-sol}
\end{equation}
so that $(\rr_z,\pp_z)$ is jointly Gaussian. Their means are thus given by
\begin{equation}
\E[\pp_z]=\pp_0,
\qquad
\E[\rr_z]=\rr_0+z\pp_0.
\label{eq:appC-mean}
\end{equation}
Its covariance blocks are
\begin{equation}
\Cov(\pp_z)=2\Deff z\,I_d,
\qquad
\Cov(\rr_z,\pp_z)=\Deff z^2\,I_d,
\qquad
\Cov(\rr_z)=\frac{2}{3}\Deff z^3\,I_d.
\label{eq:appC-covariances}
\end{equation}

To write the fundamental solution compactly, we introduce the following state vector
\begin{equation}
X=
\begin{pmatrix}
\rr\\ \pp
\end{pmatrix},
\qquad
X_0=
\begin{pmatrix}
\rr_0\\ \pp_0
\end{pmatrix}.
\label{eq:appC-X-def}
\end{equation}
Then, we can also introduce its associated transport matrix
\begin{equation}
S(z)
=
\begin{pmatrix}
I_d & z I_d\\
0 & I_d
\end{pmatrix},
\label{eq:transport-matrix}
\end{equation}
and its diffusion-generated covariance matrix as
\begin{equation}
Q(z)
=
\begin{pmatrix}
\frac{2}{3}\Deff z^3\,I_d & \Deff z^2\,I_d\\[0.4em]
\Deff z^2\,I_d & 2\Deff z\,I_d
\end{pmatrix}.
\label{eq:Q-matrix}
\end{equation}
The mean trajectory is then defined as
\begin{equation}
m(z;X_0)=S(z)X_0=
\begin{pmatrix}
r_0+zp_0\\
p_0
\end{pmatrix}.
\label{eq:mean-trajectory}
\end{equation}
The fundamental solution of \eqref{eq:minimal-fp-model} is therefore
\begin{equation}
G_z(X\,|\,X_0)
=
\frac{1}{(2\pi)^d\sqrt{\det Q(z)}}
\exp\!\left(
-\frac12
\bigl(X-S(z)X_0\bigr)^\trans
Q(z)^{-1}
\bigl(X-S(z)X_0\bigr)
\right),
\qquad z>0,
\label{eq:green-matrix}
\end{equation}
and the solution of the associated Cauchy problem is given by
\begin{equation}
\W(z,\rr,\pp)
=
\int_{\R^d}\int_{\R^d}
G_z\!\left(
\begin{pmatrix}
\rr\\ \pp
\end{pmatrix}
\Bigg|
\begin{pmatrix}
\rr_0\\ \pp_0
\end{pmatrix}
\right)
W_0(\rr_0,\pp_0)\,\dif \rr_0\,\dif \pp_0.
\label{eq:green-representation}
\end{equation}

Because all blocks of $Q(z)$ are scalar multiples of the identity, we get
\begin{equation}
\det Q(z)=\left(\frac{1}{3}\Deff^2 z^4\right)^d,
\qquad
\sqrt{\det Q(z)}=\left(\frac{\Deff z^2}{\sqrt{3}}\right)^d,
\label{eq:appC-detQ}
\end{equation}
and
\begin{equation}
Q(z)^{-1}
=
\begin{pmatrix}
\displaystyle \frac{6}{\Deff z^3}I_d & \displaystyle -\frac{3}{\Deff z^2}I_d\\[0.8em]
\displaystyle -\frac{3}{\Deff z^2}I_d & \displaystyle \frac{2}{\Deff z}I_d
\end{pmatrix}.
\label{eq:appC-invQ}
\end{equation}
It leads to the following explicit solution in scalar form
\begin{equation}
G_z(\rr,\pp\,|\,\rr_0,pp_0)
=
\left(\frac{\sqrt{3}}{2\pi \Deff z^2}\right)^d
\exp\!\left(
-\frac{|\pp-\pp_0|^2}{4\Deff z}
-
\frac{3}{\Deff z^3}
\left|
\rr-\rr_0-\frac{z}{2}(\pp+\pp_0)
\right|^2
\right).
\label{eq:green-explicit}
\end{equation}
Equivalently, introducing the two jump variables
$\delta \pp:=\pp-\pp_0$,
$\delta \rr:=\rr-\rr_0-z\pp_0$,
one may rewrite the same formula as
\begin{equation}
G_z(\rr,\pp\,|\,\rr_0,\pp_0)
=
\left(\frac{\sqrt{3}}{2\pi \Deff z^2}\right)^d
\exp\!\left(
-\frac{|\delta \pp|^2}{4\Deff z}
-
\frac{3}{\Deff z^3}
\left|
\delta \rr-\frac{z}{2}\delta \pp
\right|^2
\right).
\label{eq:green-explicit-increments}
\end{equation}
This form makes visible the two characteristic scales of the model,
\[
|\delta \pp|\sim (\Deff z)^{1/2},
\qquad
|\delta \rr-z/2\delta \pp|\sim (\Deff z^3)^{1/2},
\]
the latter being the phase-space expression of the cubic beam-spreading law derived in \cref{sec:moments}.

Suppose now that the initial phase-space density is Gaussian and writes as
\begin{equation}
W_0(\rr,\pp)
=
\frac{M}{(2\pi)^d\sqrt{\det\Sigma_0}}
\exp\!\left(
-\frac12
\bigl(X-m_0\bigr)^\trans
\Sigma_0^{-1}
\bigl(X-m_0\bigr)
\right),
\label{eq:general-gaussian-initial}
\end{equation}
with the mean $m_0=(\bar \rr_0, \bar \pp_0)^T$ and where $\Sigma_0$ is a positive definite covariance matrix of size $2d\times 2d$. Since the Green kernel is Gaussian, the solution remains Gaussian for all $z>0$ leading to
\begin{equation}
\W(z,\rr,\pp)
=
\frac{M}{(2\pi)^d\sqrt{\det\Sigma(z)}}
\exp\!\left(
-\frac12
\bigl(X-m(z)\bigr)^\trans
\Sigma(z)^{-1}
\bigl(X-m(z)\bigr)
\right),
\label{eq:general-gaussian-solution}
\end{equation}
with mean
\begin{equation}
m(z)=S(z)m_0,
\label{eq:mean-evolution}
\end{equation}
and covariance
\begin{equation}
\Sigma(z)=S(z)\Sigma_0 S(z)^\trans + Q(z).
\label{eq:covariance-evolution}
\end{equation}
Thus the deterministic part of the dynamics transports the initial covariance according to free phase-space propagation, while the random medium contributes the additional covariance $Q(z)$.

Writing
\begin{equation}
\Sigma_0=
\begin{pmatrix}
\Sigma_{\rr\rr}^{(0)} & \Sigma_{\rr\pp}^{(0)}\\
\Sigma_{\pp\rr}^{(0)} & \Sigma_{\pp\pp}^{(0)}
\end{pmatrix},
\label{eq:sigma0-block}
\end{equation}
the evolved covariance
\[
\Sigma(z)=
\begin{pmatrix}
\Sigma_{\rr\rr}(z) & \Sigma_{\rr\pp}(z)\\
\Sigma_{\pp\rr}(z) & \Sigma_{\pp\pp}(z)
\end{pmatrix}
\]
satisfies
\begin{equation}
\Sigma_{\pp\pp}(z)=\Sigma_{\pp\pp}^{(0)}+2\Deff z\,I_d,
\label{eq:sigma-pp}
\end{equation}
\begin{equation}
\Sigma_{\rr\pp}(z)=\Sigma_{\rr\pp}^{(0)}+z\,\Sigma_{\pp\pp}^{(0)}+\Deff z^2\,I_d,
\label{eq:sigma-rp}
\end{equation}
\begin{equation}
\Sigma_{\pp\rr}(z)=\Sigma_{\pp\rr}^{(0)}+z\,\Sigma_{\pp\pp}^{(0)}+\Deff z^2\,I_d,
\label{eq:sigma-pr}
\end{equation}
and
\begin{equation}
\Sigma_{\rr\rr}(z)
=
\Sigma_{\rr\rr}^{(0)}
+
z\bigl(\Sigma_{\rr\pp}^{(0)}+\Sigma_{\pp\rr}^{(0)}\bigr)
+
z^2\Sigma_{\pp\pp}^{(0)}
+
\frac{2}{3}\Deff z^3\,I_d.
\label{eq:sigma-rr}
\end{equation}
These formulas recover, at the covariance level, the triangular cascade already identified for the quadratic moments.

An interesting initial condition for the sequel is the centered isotropic Gaussian/Gauss--Schell one where the initial state writes
\begin{equation}
W_0(\rr,\pp)
=
\frac{M}{(2\pi \sigma_\rr \sigma_\pp)^d}
\exp\!\left(
-\frac{|\rr|^2}{2\sigma_\rr^2}
-
\frac{|\pp|^2}{2\sigma_\pp^2}
\right).
\label{eq:isotropic-gauss-schell}
\end{equation}
This corresponds to the case, where
$m_0=0$,
$\Sigma_{\rr\rr}^{(0)}=\sigma_\rr^2 I_d$,
$\Sigma_{\pp\pp}^{(0)}=\sigma_\pp^2 I_d$ and
$\Sigma_{\rr\pp}^{(0)}=\Sigma_{\pp\rr}^{(0)}=0$. In that case, we have
\begin{equation}
\Sigma_{\pp\pp}(z)
=
\bigl(\sigma_\pp^2+2\Deff z\bigr)I_d,
\label{eq:isotropic-sigma-pp}
\end{equation}
\begin{equation}
\Sigma_{\rr\pp}(z)
= \Sigma_{\pp\rr} (z)=
\bigl(z\sigma_\pp^2+\Deff z^2\bigr)I_d,
\label{eq:isotropic-sigma-rp}
\end{equation}
and
\begin{equation}
\Sigma_{\rr\rr}(z)
=
\left(
\sigma_\rr^2
+
z^2\sigma_\pp^2
+
\frac{2}{3}\Deff z^3
\right)I_d.
\label{eq:isotropic-sigma-rr}
\end{equation}

The Gaussian structure of the solution makes the marginals explicit. In the general Gaussian case we have
\begin{equation}
\Jm(z,\pp)
=
\frac{M}{(2\pi)^{d/2}\sqrt{\det \Sigma_{\pp\pp}(z)}}
\exp\!\left(
-\frac12
\bigl(\pp-\bar \pp(z)\bigr)^\trans
\Sigma_{\pp\pp}(z)^{-1}
\bigl(\pp-\bar \pp(z)\bigr)
\right),
\label{eq:J-general}
\end{equation}
and
\begin{equation}
\I(z,\rr)
=
\frac{M}{(2\pi)^{d/2}\sqrt{\det \Sigma_{\rr\rr}(z)}}
\exp\!\left(
-\frac12
\bigl(\rr-\bar \rr(z)\bigr)^\trans
\Sigma_{\rr\rr}(z)^{-1}
\bigl(\rr-\bar \rr(z)\bigr)
\right),
\label{eq:I-general}
\end{equation}
where $\bar \pp(z)$ and $\bar \rr(z)$ denote the momentum and spatial components of the mean vector $m(z)$. In the centered isotropic case \eqref{eq:isotropic-gauss-schell}, these further reduce to
\begin{equation}
\Jm(z,\pp)
=
\frac{M}
{\bigl(2\pi(\sigma_\pp^2+2\Deff z)\bigr)^{d/2}}
\exp\!\left(
-\frac{|\pp|^2}{2(\sigma_\pp^2+2\Deff z)}
\right),
\label{eq:J-isotropic}
\end{equation}
and
\begin{equation}
\I(z,\rr)
=
\frac{M}
{\left(2\pi\left(\sigma_\rr^2+z^2\sigma_\pp^2+\frac{2}{3}\Deff z^3\right)\right)^{d/2}}
\exp\!\left(
-\frac{|\rr|^2}
{2\left(\sigma_\rr^2+z^2\sigma_\pp^2+\frac{2}{3}\Deff z^3\right)}
\right).
\label{eq:I-isotropic}
\end{equation}
The three contributions to the beam width are fully separated. From the initial spatial width $\sigma_\rr^2$, it undergoes a ballistic broadening $z^2\sigma_\pp^2$ due to the initial angular spread, and a cumulative broadening $\frac{2}{3}\Deff z^3$ induced by the random medium.

To conclude this part, these formulas also provide an independent recovery of the moment laws derived in \cref{sec:moments}. Indeed, in the isotropic centered case, we get
\begin{equation}
P_2(z)
=
\Tr\Sigma_{\pp\pp}(z)
=
d\sigma_\pp^2+2d\Deff z,
\label{eq:P2-from-gaussian}
\end{equation}
and
\begin{equation}
R_2(z)
=
\Tr\Sigma_{\rr\rr}(z)
=
d\sigma_\rr^2+d\sigma_\pp^2 z^2+\frac{2d}{3}\Deff z^3.
\label{eq:R2-from-gaussian}
\end{equation}
In particular, for $d=2$, we obtain
\begin{equation}
P_2(z)=2\sigma_\pp^2+4\Deff z,
\qquad
R_2(z)=2\sigma_\rr^2+2\sigma_\pp^2 z^2+\frac{4}{3}\Deff z^3,
\label{eq:moment-laws-d2-again}
\end{equation}
in agreement with the laws obtained in the previous section.


\section{Numerical experiments}
\label{sec:num-exp}

The theoretical framework developed above has been written for an arbitrary transverse dimension $d$,
with $d=2$ corresponding to the standard paraxial beam-propagation setting. The numerical
experiments reported in this section are performed in the one-dimensional case, i.e., $d=1$ and $(r,p)\in \R \times \R$. This choice is deliberate. It is sufficient to test the coupling between ballistic transport in $r$, redistribution or diffusion in $p$, the quadratic-moment cascade, and the kinetic-to-diffusive reduction, while keeping the numerical tests reproducible. Indeed, in the homogeneous isotropic setting, the extension to higher transverse dimension is only tensorial. The one-dimensional tests thus already probe the essential closure mechanism.

\subsection{Discretization and validation strategy}
\label{sec:numerical-discretization}

This subsection summarizes the numerical framework used for the one-dimensional transverse
phase-space models considered in the paper. The focus is on the minimal Fokker--Planck model
\begin{equation}
\partial_z \W(z,r,p)
+
p\,\partial_r \W(z,r,p)
=
\Deff\,\partial_{pp}\W(z,r,p),
\label{eq:num-fp}
\end{equation}
and on the nonlocal kinetic reference model
\begin{equation}
\partial_z \W(z,r,p)
+
p\,\partial_r \W(z,r,p)
=
\int_{\R}
K(q)\left[\W(z,r,p+q)-\W(z,r,p)\right]\dif q.
\label{eq:num-kin}
\end{equation}
In Equation~\eqref{eq:num-fp}, we defined $\Deff$ as 
\begin{equation}
\Deff
=
\frac12\int_{\R}q^2K(q)\,\dif q,
\label{eq:num-deff-1d}
\end{equation}
The aim of the numerical strategy is to preserve the transport structure in $(r,p)$, while resolving accurately the diffusion or jump dynamics in momentum, and compare the numerical solutions with the analytical benchmarks derived in the above sections.

We work in the following transverse computational domain
$\Omega=[-R_{\max},R_{\max}]\times[-P_{\max},P_{\max}]$. A standard
uniform grid is used for the discretization
\begin{equation}
r_i=-R_{\max}+i\,\Delta r,
\qquad
p_j=-P_{\max}+j\,\Delta p,
\label{eq:num-grid}
\end{equation}
with
\begin{equation}
\Delta r=\frac{2R_{\max}}{N_r-1},
\qquad
\Delta p=\frac{2P_{\max}}{N_p-1}.
\label{eq:num-grid-steps}
\end{equation}
The discrete phase-space density is denoted by
\begin{equation}
W^n_{i,j}\simeq \W(z_n,r_i,p_j),
\qquad
z_n=n\,\Delta z.
\label{eq:num-discrete-density}
\end{equation}
The computational domains are chosen large enough so that the solution remains negligible near the boundaries over the propagation range of interest. In other cases, one can add apodization layers at the boundaries.

For the minimal model \eqref{eq:num-fp}, the propagation is split into a transport step
\begin{equation}
\partial_z\W+p\,\partial_r\W=0
\label{eq:num-transport-substep}
\end{equation}
and a diffusion step
\begin{equation}
\partial_z\W=\Deff\,\partial_{pp}\W.
\label{eq:num-diffusion-substep}
\end{equation}
At first order, this gives
\begin{equation}
W^{n+\frac12}=\mathcal{T}_{\Delta z}W^n,
\qquad
W^{n+1}=\mathcal{D}_{\Delta z}W^{n+\frac12}.
\label{eq:num-lie-splitting}
\end{equation}
For second order accuracy, one can use a Strang splitting~\cite{macnamara2017operator} leading to
\begin{equation}
W^{n+1}
=
\mathcal{D}_{\Delta z/2}
\mathcal{T}_{\Delta z}
\mathcal{D}_{\Delta z/2}
W^n.
\label{eq:num-strang-splitting}
\end{equation}

The transport step is treated semi-Lagrangianly along exact characteristics as
\begin{equation}
\W(z+\Delta z,r,p)=\W(z,r-\Delta z\,p,p).
\label{eq:num-exact-characteristics}
\end{equation}
At the discrete level, this becomes
\begin{equation}
W^{n+\frac12}_{i,j}
=
\mathcal{I}_r\left[W^n(\cdot,p_j)\right]
\left(r_i-\Delta z\,p_j\right),
\label{eq:num-transport-update}
\end{equation}
where $\mathcal{I}_r$ denotes an interpolation operator in the spatial variable. Here, we use a linear interpolation which is sufficient when $d=1$. This semi-Lagrangian treatment respects the exact characteristic structure and avoids a CFL restriction associated with explicit Eulerian advection schemes.

For the diffusion step, each spatial slice $W(r_i,\cdot)$ is advanced directly in discrete Fourier variables with respect to the momentum grid. Let
$\widehat{W}^{\,n}_{i,\ell}$ denote the discrete Fourier transform of $W^n(r_i,\cdot)$, and $k_\ell$ the associated
discrete Fourier wave numbers. Since the diffusion operator is diagonal in the spectral domain, each mode is updated exactly over one step according to
\begin{equation}
\widehat{W}^{\,n+1}_{i,\ell}
=
\exp\left(-\Deff\,\Delta z\,k_\ell^2\right)
\widehat{W}^{\,n+\frac12}_{i,\ell}.
\label{eq:num-diffusion-update-fourier}
\end{equation}
Going back to physical momentum space then gives
\begin{equation}
W^{n+1}(r_i,\cdot)
=
\mathcal{F}_p^{-1}
\left[
\exp\left(-\Deff\,\Delta z\,k_\ell^2\right)
\mathcal{F}_p\left(W^{n+\frac12}(r_i,\cdot)\right)
\right].
\label{eq:num-diffusion-update}
\end{equation}
Thus the diffusion step is implemented by exact mode-by-mode multiplication in the discrete spectral domain associated to the momentum grid. Since $P_{\max}$ is chosen large enough, the solution and the kernel are negligible near the boundaries, avoiding spurious reflections inside the computational domain. In addition, the overall complexity of the method is $\mathcal{O}(NN_p\log N_p)$, with $N$ the total number of propagation steps.

The same framework is used for the nonlocal kinetic model, the only change being the momentum substep. For a homogeneous kernel, the momentum operator reads
\begin{equation}
\Qkin[\W](p)
=
\int_{\R}
K(q)\left[\W(p+q)-\W(p)\right]\dif q.
\label{eq:num-kinetic-operator-1d}
\end{equation}
If, in addition, the kernel is symmetric, $K(q)=K(-q)$, the gain term may be written as 
\begin{equation}
\Qkin[\W]
=
K *_p \W - \lambda \W,
\qquad
\lambda=\int_{\R}K(q)\,\dif q,
\label{eq:num-kinetic-convolution}
\end{equation}
where $*_p$ denotes convolution in $p$, which is diagonal in the Fourier domain. In the spectral domain, it thus writes as
\begin{equation}
\partial_z\widehat{W}
=
\left(\widehat{K}(k_p)-\lambda\right)\widehat{W}.
\label{eq:num-kinetic-fourier-generator}
\end{equation}
The corresponding update is therefore
\begin{equation}
W^{n+1}(r_i,\cdot)
=
\mathcal{F}_p^{-1}
\left[
\exp\left(\left(\widehat{K}(k_p)-\lambda\right)\Delta z\right)
\mathcal{F}_p\left(W^{n+\frac12}(r_i,\cdot)\right)
\right].
\label{eq:num-kinetic-update}
\end{equation}
At the discrete level, the same structure is retained, with $\lambda$ replaced by the discrete mass of the kernel on the momentum grid. This makes the comparison between the nonlocal kinetic model and its Fokker--Planck approximation direct. Indeed both are solved by the same transport-splitting strategy and differ only through the Fourier multiplier used in the momentum space.

At each propagation step we compute the discrete analogues of the main observables,
\begin{align}
M^n
&=
\Delta r\,\Delta p
\sum_{i,j} W^n_{i,j},
\label{eq:num-mass-discrete}
\\
P_2^n
&=
\Delta r\,\Delta p
\sum_{i,j}p_j^2 W^n_{i,j},
\label{eq:num-P2-discrete}
\\
C^n
&=
\Delta r\,\Delta p
\sum_{i,j} r_i p_j\, W^n_{i,j},
\label{eq:num-C-discrete}
\\
R_2^n
&=
\Delta r\,\Delta p
\sum_{i,j}r_i^2 W^n_{i,j}.
\label{eq:num-R2-discrete}
\end{align}
We also compute the spatial and angular marginals
\begin{equation}
\I^n(r_i)
=
\Delta p\sum_j W^n_{i,j},
\qquad
\Jm^n(p_j)
=
\Delta r\sum_i W^n_{i,j}.
\label{eq:num-marginals-discrete}
\end{equation}
Mass conservation is monitored through
\begin{equation}
\varepsilon_M^n
=
\frac{|M^n-M^0|}{M^0},
\label{eq:num-mass-error}
\end{equation}
and the occurrence of negative numerical undershoots is monitored through
\begin{equation}
m_{\min}^n=\min_{i,j}W^n_{i,j}.
\label{eq:num-min-value}
\end{equation}
In the benchmark regimes considered here, such undershoots remain negligible for the observables of interest. Therefore, no positivity-preserving limiter is introduced. Furthermore convergence is assessed by systematic refinement of $\Delta z$, $N_r$, $N_p$, and the truncation parameters $R_{\max}$ and $P_{\max}$. When an exact solution is available, we also monitor relative profile errors, for instance
\begin{equation}
\varepsilon_{\I}(z_n)
=
\frac{\left\|\I^n-\I_{\mathrm{exact}}(z_n,\cdot)\right\|_{\ell^2}}
{\left\|\I_{\mathrm{exact}}(z_n,\cdot)\right\|_{\ell^2}},
\label{eq:num-profile-error}
\end{equation}
and similarly for $\Jm$, with $\ell^2$ the discrete $L^2$ norm. 

For the one-dimensional transverse benchmark, the general $d$-dimensional moment identities
reduce to
\begin{align}
P_2(z)
&=
P_{2,0}+2\Deff Mz,
\label{eq:num-P2-exact-1d}
\\
C(z)
&=
C_0+P_{2,0}z+\Deff Mz^2,
\label{eq:num-C-exact-1d}
\\
R_2(z)
&=
R_{2,0}+2C_0z+P_{2,0}z^2+\frac{2}{3}\Deff Mz^3.
\label{eq:num-R2-exact-1d}
\end{align}
For centered normalized Gaussian initial data with $C_0=0$, these formulas are the reference curves used in the numerical validation.



\subsection{Range of validity and limitations of the minimal model}
\label{sec:validity-limits}

First, let us study the range of validity of the minimal Fokker--Planck model~\eqref{eq:num-fp}. Indeed the latter is as a reduced closure, not as a universal first-principles model for wave propagation in random media. 

\begin{figure}[htbp]
    \centering
    \includegraphics[trim={0.2cm 0.2cm 0.2cm 0.8cm},clip,width=0.78\textwidth]{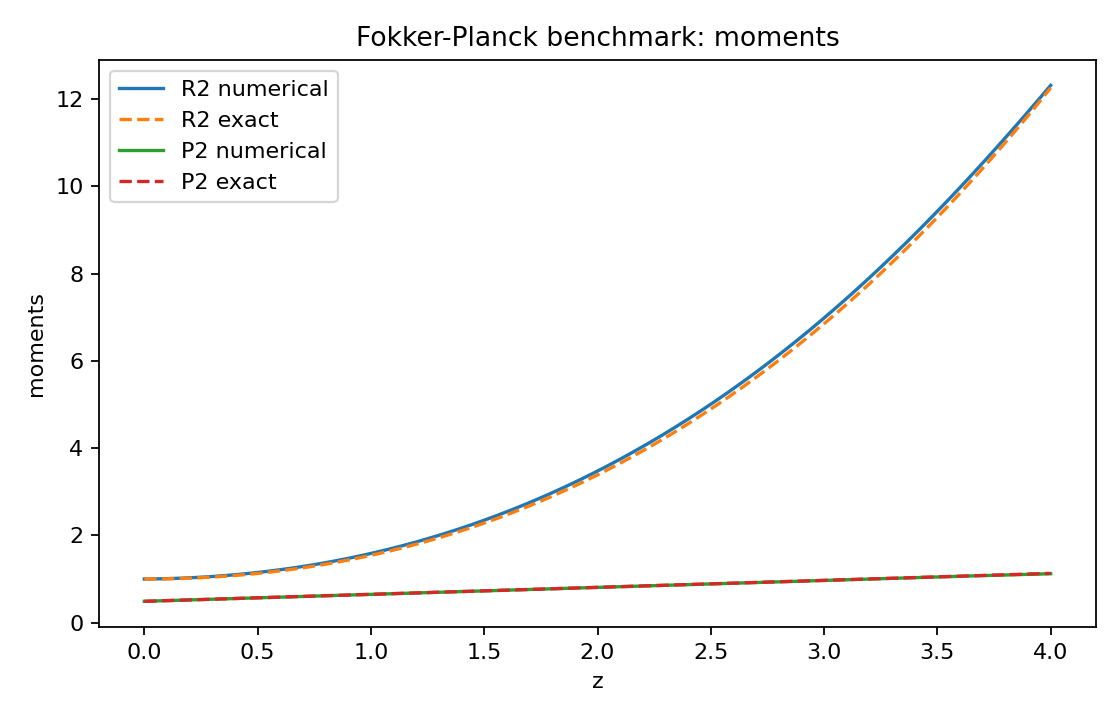}
    \caption{
    Validation of the one-dimensional transverse reduction of the minimal Fokker--Planck computational model against the explicit Gaussian solution. }
    \label{fig:fp-benchmark}
\end{figure}

Our first goal is to validate the numerical method derived in the previous section. In Figure~\ref{fig:fp-benchmark}, we compare the numerical evolution, i.e. for the minimal model solver, of the quadratic moments with the exact one-dimensional Gaussian formulas of~\eqref{eq:num-P2-exact-1d}--\eqref{eq:num-R2-exact-1d}. 

The results provided by the numerical solver are in good agreement with the theoretical benchmark, validating the proposed approach.



Because the minimal model arises from a second-order expansion in momentum transfer, its natural benchmark is not the exact random wave equation itself, but the intermediate nonlocal kinetic closure~\eqref{eq:num-kin}. Indeed, this equation preserves the jump structure of the effective phase-space redistribution and therefore retains the information that is lost in the local diffusive approximation. In particular, it remains sensitive to finite momentum transfers and to the detailed shape of the redistribution kernel.



First, let us consider a centered Gaussian jump kernels, which is defined as
\begin{equation}
K_\eps(q)=A_\eps \exp\left(-\frac{q^2}{2\eps^2}\right),
\label{eq:narrow-kernel-validity}
\end{equation}
with normalization chosen so that the effective diffusion coefficient remains fixed while the width $\eps$ is varied. 

\begin{figure}[htbp]
    \centering
    \begin{minipage}{0.48\textwidth}
        \centering
        \includegraphics[trim={0.2cm 0.2cm 0.2cm 0.8cm},clip,width=\textwidth]{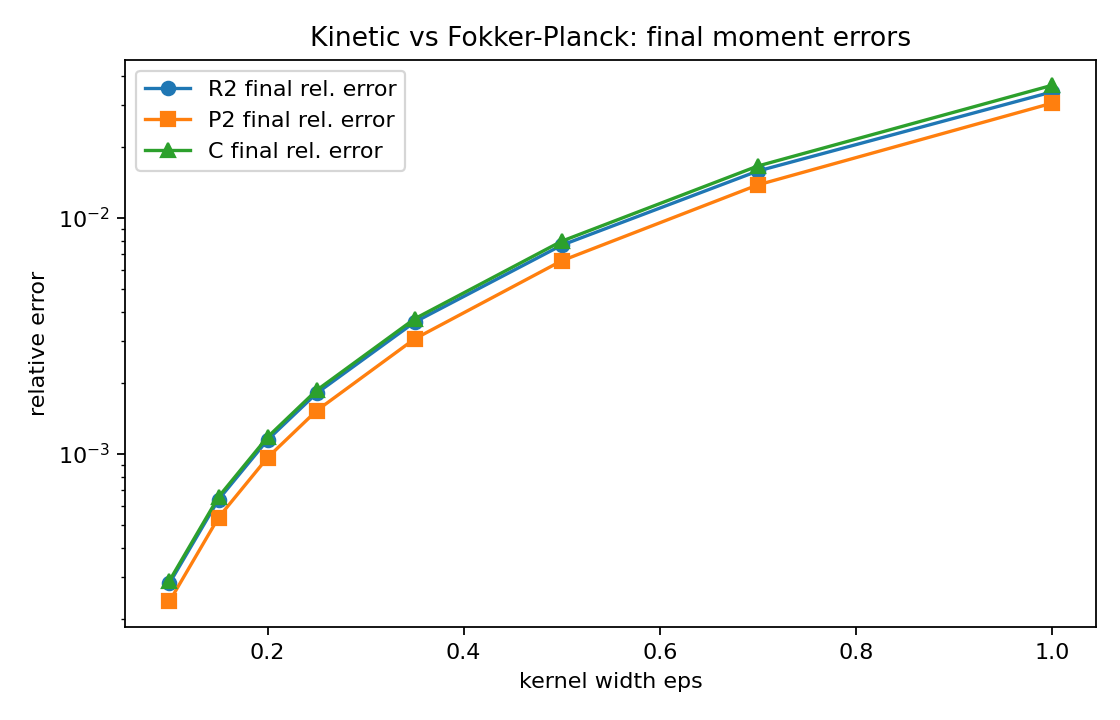}
    \end{minipage}
    \hfill
    \begin{minipage}{0.48\textwidth}
        \centering
        \includegraphics[trim={0.2cm 0.2cm 0.2cm 0.8cm},clip,width=\textwidth]{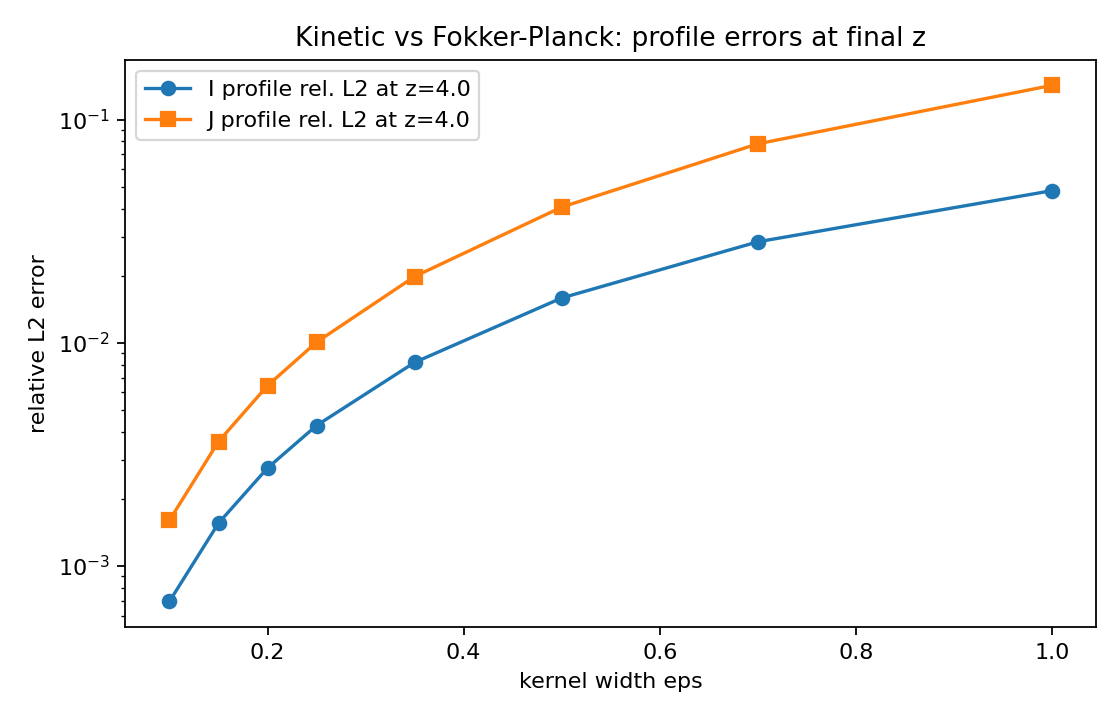}
    \end{minipage}
    \caption{
    One-dimensional transverse kinetic-to-diffusive benchmark for Gaussian momentum-transfer kernels of increasing width $\eps$. Left: final relative errors on the quadratic moments. Right: final relative errors on the spatial and angular marginals.
    }
    \label{fig:gaussian-validity}
\end{figure}

In Figure~\ref{fig:gaussian-validity} we plot the relative error for the different observables with the kernel width. For narrow Gaussian kernels, the agreement between the nonlocal kinetic evolution and the Fokker--Planck approximation is excellent, both at the level of quadratic moments and at the level of the spatial and angular marginals. As the kernel width increases, the discrepancy grows monotonically, as expected. Moreover, the first observable to depart significantly from the kinetic reference is the angular marginal, whereas the low-order global moments remain comparatively robust over a wider range of kernel widths. Indeed, as shown in the theoretical analysis, i.e. cascade behaviour, the angular profile is the most sensitive to the nonlocal momentum redistribution, while the integrated beam-width observables react only at a later stage.

\begin{figure}[htbp]
    \centering
    \begin{minipage}{0.48\textwidth}
        \centering
        \includegraphics[trim={0.2cm 0.2cm 0.2cm 0.8cm},clip,width=\textwidth]{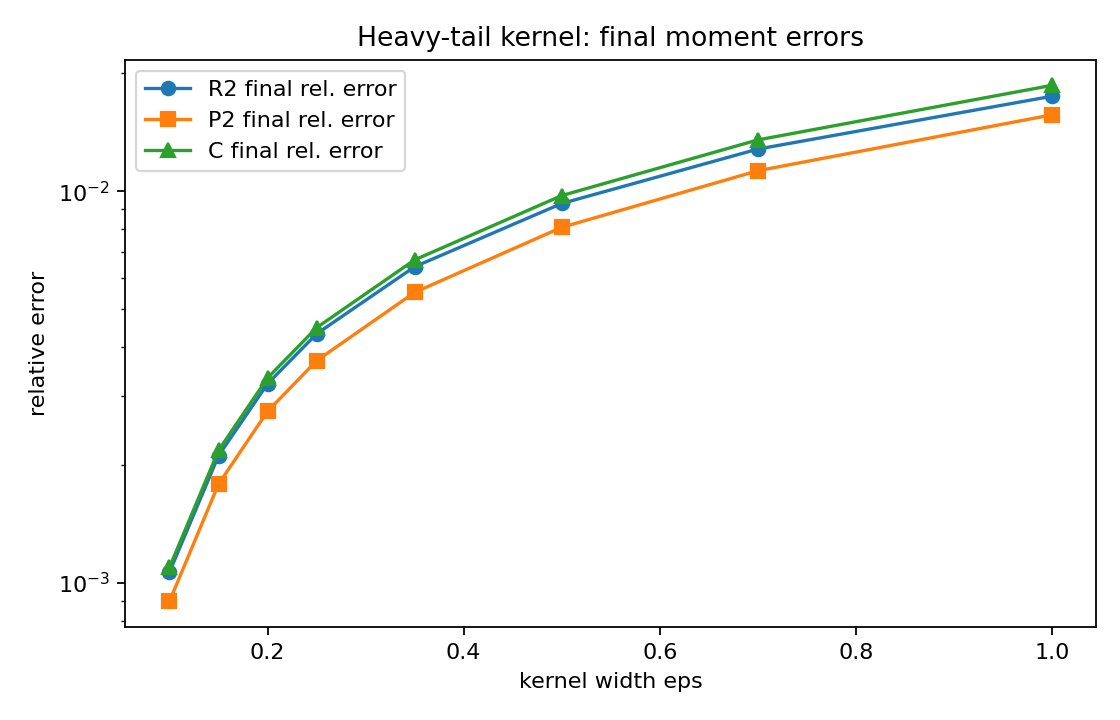}
    \end{minipage}
    \hfill
    \begin{minipage}{0.48\textwidth}
        \centering
        \includegraphics[trim={0.2cm 0.2cm 0.2cm 0.8cm},clip,width=\textwidth]{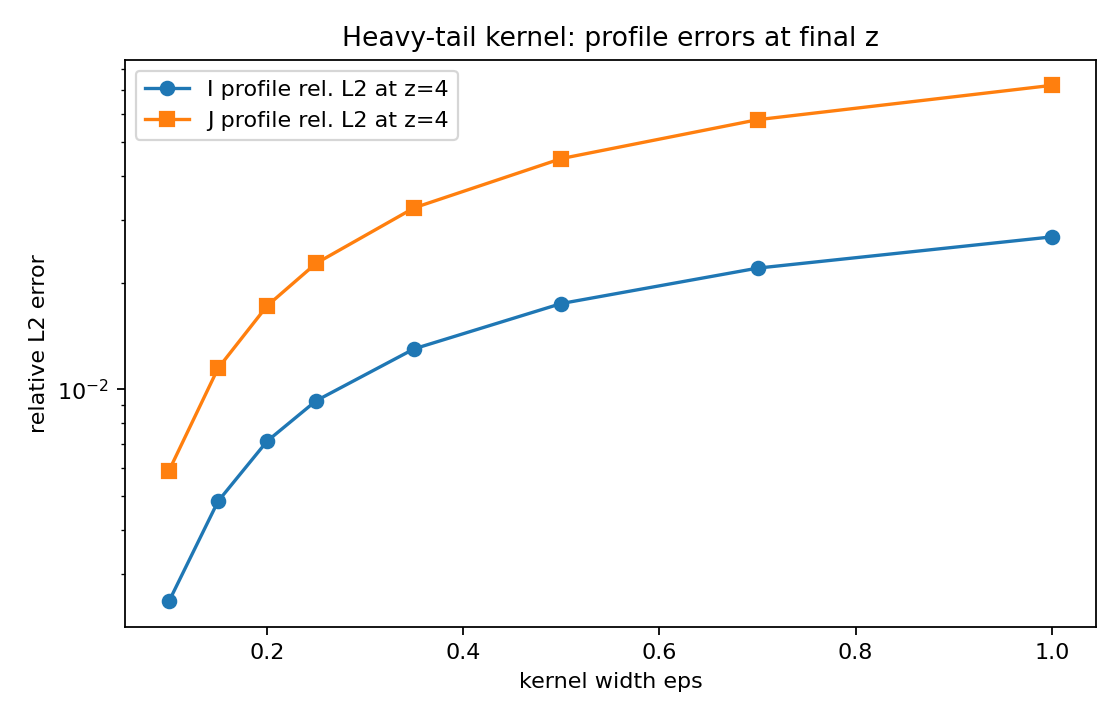}
    \end{minipage}
    \caption{
    Influence of kernel shape on the validity of the diffusive closure in the one-dimensional transverse benchmark. Left: final relative errors on the quadratic moments. Right: final relative errors on the spatial and angular marginals.
    }
    \label{fig:heavy-tail-validity}
\end{figure}

To test the influence of kernel shape beyond second-moment matching, we next compare the Fokker--Planck approximation with redistribution laws that are broader than Gaussian narrow-jump kernels while remaining exponentially regularized. To this end, we define
\begin{equation}
K_\eps(q)=A_\eps \frac{\exp(-|q|/\eps)}{1+|q|^m},
\qquad m=4,
\label{eq:broader-tail-kernel}
\end{equation}
which is normalized so that the effective diffusion coefficient matches that of the local model. As before, the corresponding results are displayed in Figure.~\ref{fig:heavy-tail-validity}. They confirm that the validity of the diffusive closure is controlled not only by the nominal width of the kernel, but also
by its detailed shape. In particular, kernels with broader nonlocal redistribution features may produce larger deviations in angular observables even when the second jump moment is held fixed. Thus, second-moment matching alone does not fully determine the accuracy of the local closure.

Taken together, these two numerical tests show that the local diffusive closure is an asymptotic version of the general nonlocal kinetic model. As expected from our assumptions, in the narrow-jump regime, it reproduces the kinetic dynamics with excellent accuracy. Away from that regime, the discrepancy grows in a controlled
and interpretable way, first in angular observables and then in the lower-order integrated moments.

\subsection{A first atmospheric specialization}
\label{sec:atmospheric-specialization}

The purpose of this section is not to derive a complete transport theory of atmospheric turbulence, but to show how the closure hierarchy developed above can be connected to a physically meaningful parameterization of refractive-index fluctuations. The atmospheric content is introduced through
an effective momentum-transfer law, and hence through the coefficient $\Deff$ appearing in the minimal Fokker--Planck model.

To relate $K$ from~\eqref{eq:num-kin} to refractive-index fluctuations, we introduce an effective transverse spectral density $\Phi_{\mathrm{eff}}(q)$, i.e. the turbulence spectrum~\cite{fabbro2013scintillation,morel2025ionospheric}, assumed nonnegative and radial. We can thus rewrite $K$ as
\begin{equation}
K(q)=\Gamma_{\atm}\,\Phi_{\mathrm{eff}}(q),
\label{eq:k-via-phi}
\end{equation}
where $\Gamma_{\atm}>0$ is an effective proportionality factor collecting the dimensional and
asymptotic constants required to map the medium statistics to the momentum-transfer law. The
corresponding diffusion coefficient is then
\begin{equation}
\Deff
=
\frac{\Gamma_{\atm}}{2d}
\int_{\R^d}
|q|^2 \Phi_{\mathrm{eff}}(q)\,\dif q,
\label{eq:deff-via-phi}
\end{equation}
or, in dimension $d=2$ under radial symmetry,
\begin{equation}
\Deff
=
\frac{\pi \Gamma_{\atm}}{2}
\int_0^\infty
\rho^3 \Phi_{\mathrm{eff}}(\rho)\,\dif \rho.
\label{eq:deff-via-phi-radial}
\end{equation}
This representation is deliberately flexible: the closure hierarchy remains unchanged, and the
atmospheric content enters only through the effective spectral shape of the redistribution law.

For a first atmospheric specialization, we adopt a regularized von K\'arm\'an-type spectrum~\cite{von1951statistical}
\begin{equation}
\Phi_{\mathrm{eff}}(\rho)
=
C_n^2
\frac{\exp(-\rho^2/\kappa_m^2)}
{\left(\rho^2+\kappa_0^2\right)^{11/6}},
\label{eq:phi-eff-vonkarman}
\end{equation}
with
\begin{equation}
\kappa_0=\frac{1}{L_0},
\qquad
\kappa_m=\frac{5.92}{l_0},
\label{eq:kappa-def}
\end{equation}
where $C_n^2$ is the refractive-index structure parameter, $L_0$ the outer scale, and $l_0$ the inner scale. The exponent $11/6$ reflects the usual Kolmogorov-type inertial scaling in the reduced radial representation. The latter is widely used in tropospheric or ionospheric propagation model~\cite{fabbro2013scintillation,darchy2024theoretical,morel2024sar,morel2025ionospheric}. Substituting~\eqref{eq:phi-eff-vonkarman} into~\eqref{eq:k-via-phi} leads to
\begin{equation}
K(\rho)
=
\Gamma_{\atm}\,C_n^2
\frac{\exp(-\rho^2/\kappa_m^2)}
{\left(\rho^2+\kappa_0^2\right)^{11/6}},
\label{eq:k-radial-vonkarman}
\end{equation}
and therefore
\begin{equation}
\Deff
=
\frac{\pi\Gamma_{\atm}C_n^2}{2}
\int_0^\infty
\rho^3
\frac{\exp(-\rho^2/\kappa_m^2)}
{\left(\rho^2+\kappa_0^2\right)^{11/6}}
\,\dif \rho.
\label{eq:deff-vonkarman}
\end{equation}
Equation \eqref{eq:deff-vonkarman} provides a first explicit link between the minimal Wigner
transport model and standard atmospheric turbulence parameters.



First, since the integral is linear in $C_n^2$, one has
\begin{equation}
\Deff \propto C_n^2
\label{eq:deff-Cn2-scaling}
\end{equation}
for fixed $L_0$, $l_0$, and $\Gamma_{\atm}$. In the one-dimensional transverse scenario, the turbulence-induced contributions to the moments thus satisfy
\begin{equation}
P_2(z)-P_{2,\mathrm{free}}(z)\simeq 2\,\Deff z,
\qquad
R_2(z)-R_{2,\mathrm{free}}(z)\simeq \frac{2}{3}\Deff z^3.
\label{eq:atmo-1d-scaling}
\end{equation}
In the full $d$-dimensional isotropic theory, these pre-factors are replaced by $2d$ and $2d/3$, respectively. Increasing $C_n^2$ should therefore steepen the growth of the angular variance, reinforce the cubic contribution to beam spreading, and broaden both spatial and angular marginals at fixed propagation distance.

\begin{figure}[htbp]
    \centering
    \begin{minipage}{0.48\textwidth}
        \centering
        \includegraphics[trim={0.2cm 0.2cm 0.2cm 0.8cm},clip,width=\textwidth]{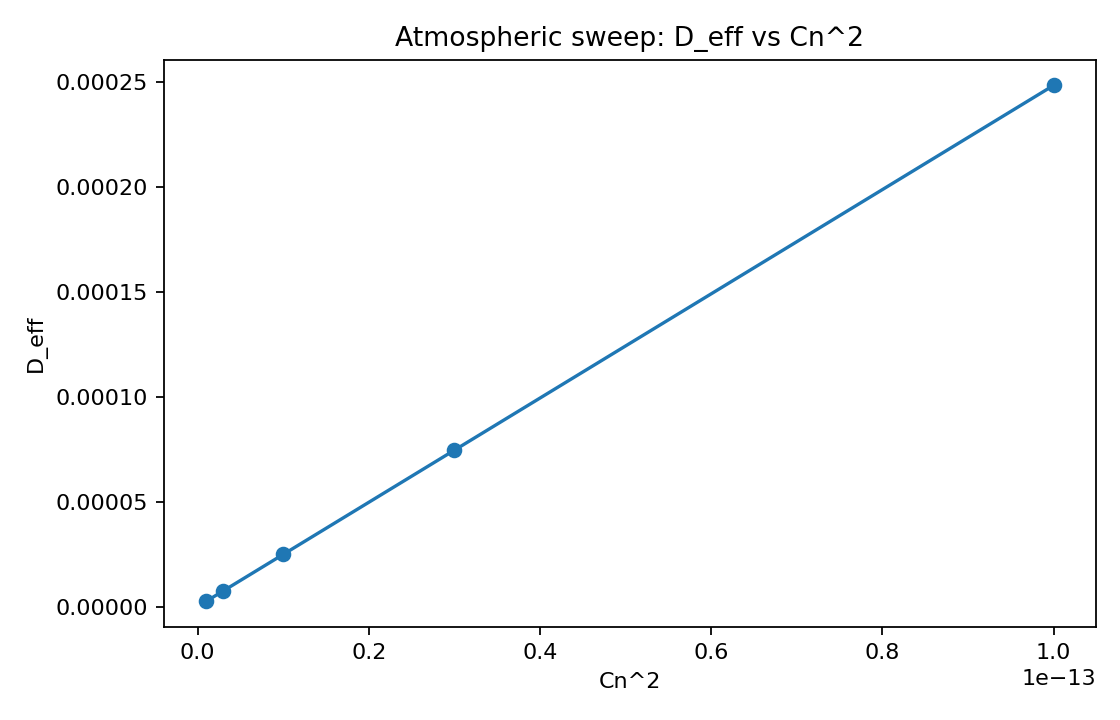}
    \end{minipage}
    \hfill
    \begin{minipage}{0.48\textwidth}
        \centering
        \includegraphics[trim={0.2cm 0.2cm 0.2cm 0.8cm},clip,width=\textwidth]{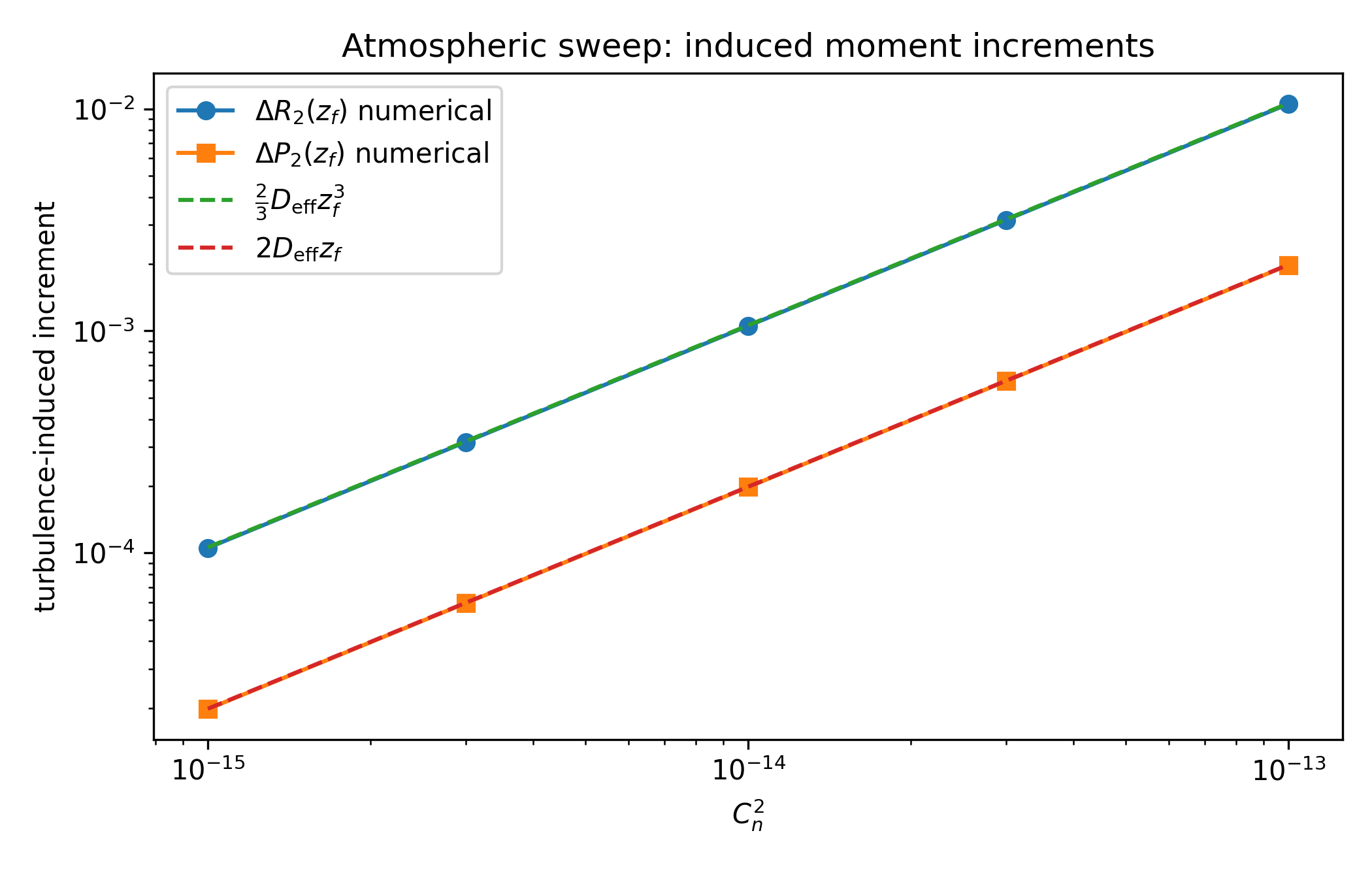}
    \end{minipage}
    \caption{
    First atmospheric parameter study for a Von-Karman type turbulence spectrum. Left: effective diffusion coefficient $\Deff$ as a function of the turbulence strength parameter $C_n^2$. Right: turbulence-induced increments of the one-dimensional quadratic moments at the final propagation distance.
    }
    \label{fig:atmo-first-sweep}
\end{figure}

Therefore, for this first numerical test, we propose to test different values of $C_n ^2$. In particular, in Figure~\ref{fig:atmo-first-sweep}, we plot both the evolution of the diffusion constant with the latter, left panel, and the turbulence increment of $P_2$ and $R_2$, right panel. The latter are defined as 
\begin{equation}
\Delta P_2(z_f)=P_2(z_f)-P_{2,\mathrm{free}}(z_f),
\qquad
\Delta R_2(z_f)=R_2(z_f)-R_{2,\mathrm{free}}(z_f).
\label{eq:atmo-moment-increments}
\end{equation}
Indeed, this allows to highlight the turbulence induced effect without initial and ballistic contributions.

From Figure~\ref{fig:atmo-first-sweep}, we can see that we accurately retrieve the the expected linear dependence on $C_n^2$. Nonetheless, the objective of this atmospheric specialization should remain clear. Even if, it establishes that the general closure hierarchy developed in this paper admits a physically coherent atmospheric realization in which the effective kinetic kernel, it has some limitations.  In particular, the exact relation between the medium covariance and the constant $\Gamma_{\atm}$ is not derived here. In addition, the numerical tests remain one-dimensional, the present model remains isotropic and local in momentum at the diffusive level. Therefore anisotropic or strongly nonlocal atmospheric transfer effects are not included.

\section{Conclusion}
\label{sec:conclusion}

We have developed a Wigner-based phase-space framework for the paraxial propagation in random media. Our main result is that volumetric randomness can be incorporated into Wigner transport in a controlled hierarchical way, with each reduction and its assumptions kept explicit.

Starting from the random paraxial wave equation, we identified the ensemble-averaged Wigner distribution as the relevant second-order state variable. The exact averaged equation contains a closure defect, expressed through a mixed field--medium correlation. This led us to propose a second ordure closure yielding a non-local kinetic closure equation. From the latter, assuming a small-angle propagation and smooth variations in the momentum space, we proposed a minimal Fokker--Planck closure. In this case we further derived the quadratic-moment system, obtained the cubic-in-distance contribution to beam spreading, and computed explicit Gaussian and Gauss--Schell propagation formulas.

In addition numerical experiments have been performed. They support this hierarchical closure. First, we have validated the phase-space solver against the explicit Gaussian benchmarks. Then we showed that the Fokker--Planck approximation agrees with the nonlocal kinetic model in the narrow-jump regime, and that the discrepancy increases when the redistribution kernel becomes broader or departs from a narrow Gaussian form, in line with the assumptions we made. Finally, a first atmospheric parameter study further shows a coherent trends in the effective diffusion coefficient and in the turbulence-induced beam-spreading contributions.

Nonetheless, the present work has some limitations. It only describes second-order averaged quantities, and thus does not provide closure for fourth-order statistics, such as scintillation, or realization-level speckle.  Likewise, the atmospheric specialization proposed here is a first effective parameterization, not a full first-principles derivation of the kinetic kernel from Kolmogorov or von K\'arm\'an turbulence.

Natural extensions include adding the deterministic refractive drift term, retaining the nonlocal kinetic level when finite momentum jumps are important, and developing higher-order closures for scintillation and speckle. We are also working on a comparison with the multiple phase screen method widely used in the electromagnetic community.


\appendix

\section{Derivation of the exact Wigner equation and closure defect}
\label{app:wigner-derivation}

We derive the exact evolution equation satisfied by the realization-dependent Wigner distribution associated with
\begin{equation}
2ik_0\,\partial_z\psi(z,\rr)+\Delta_{\rr}\psi(z,\rr)+2k_0^2\eta(z,\rr)\psi(z,\rr)=0.
\label{eq:app-paraxial}
\end{equation}
For a fixed realization, we then define 
\begin{equation}
\Gamma_\psi(z,\rr,\yy)
:=
\psi\!\left(z,\rr+\frac{\yy}{2}\right)
\psi^*\!\left(z,\rr-\frac{\yy}{2}\right),
\label{eq:app-gamma-def}
\end{equation}
and its associated Wigner transform
\begin{equation}
\Wpsi(z,\rr,\pp)
=
\left(\frac{k_0}{2\pi}\right)^d
\int_{\R^d}
\Gamma_\psi(z,\rr,\yy)\,
e^{-ik_0\pp\cdot \yy}\,\dif \yy.
\label{eq:app-wigner-def}
\end{equation}

Writing $\rr_\pm=\rr\pm \yy/2$, differentiation of $\Gamma_\psi$ using \eqref{eq:app-paraxial} and its complex conjugate leads to
\begin{equation}
\partial_z\Gamma_\psi
=
\frac{i}{2k_0}
\bigl(\Delta_{\rr_+}-\Delta_{\rr_-}\bigr)\Gamma_\psi
+
ik_0\bigl(\eta(z,\rr_+)-\eta(z,\rr_-)\bigr)\Gamma_\psi.
\label{eq:app-gamma-evol}
\end{equation}
Then, since
\[
\nabla_{\rr_+}=\frac12\nabla_{\rr}+\nabla_{\yy},
\qquad
\nabla_{\rr_-}=\frac12\nabla_{\rr}-\nabla_{\yy},
\]
we directly get
\[
\Delta_{\rr_+}-\Delta_{\rr_-}=
\left(\frac12\nabla_{\rr}+\nabla_{\yy}\right)\cdot\left(\frac12\nabla_{\rr}+\nabla_{\yy}\right)-
\left(\frac12\nabla_{\rr}-\nabla_{\yy}\right)\cdot\left(\frac12\nabla_{\rr}-\nabla_{\yy}\right)
=2\nabla_{\rr}\cdot\nabla_{\yy}.
\]
Hence
\begin{equation}
\partial_z\Gamma_\psi
=
\frac{i}{k_0}\nabla_{\rr}\cdot\nabla_{\yy}\Gamma_\psi
+
ik_0
\bigl(\eta(z,\rr+\yy/2)-\eta(z,\rr-\yy/2)\bigr)\Gamma_\psi.
\label{eq:app-gamma-evol-final}
\end{equation}

Applying the Wigner transform then yields
\begin{equation}
\partial_z\Wpsi(z,\rr,\pp)
+
\pp\cdot\nabla_{\rr}\Wpsi(z,\rr,\pp)
=
\Theta_\eta[\Wpsi](z,\rr,\pp),
\label{eq:app-exact-wigner}
\end{equation}
with
\begin{equation}
\Theta_\eta[\Wpsi](z,\rr,\pp)
=
ik_0
\left(\frac{k_0}{2\pi}\right)^d
\int_{\R^d}
\bigl(\eta(z,\rr+\yy/2)-\eta(z,\rr-\yy/2)\bigr)
\Gamma_\psi(z,\rr,\yy)\,
e^{-ik_0\pp\cdot \yy}\,\dif y.
\label{eq:app-theta-real}
\end{equation}
or in its equivalent Fourier form
\begin{equation}
\Theta_\eta[\Wpsi](z,\rr,\pp)
=
\frac{ik_0}{(2\pi)^d}
\int_{\R^d}
\widehat{\eta}(z,\mathbf{q})e^{i\mathbf{q}\cdot \rr}
\left[
\Wpsi\!\left(z,\rr,\pp-\frac{\mathbf{q}}{2k_0}\right)
-
\Wpsi\!\left(z,\rr,\pp+\frac{\mathbf{q}}{2k_0}\right)
\right]
\,\dif \mathbf{q},
\label{eq:app-theta-fourier}
\end{equation}
where 
\begin{equation}
\eta(z,r)=\frac{1}{(2\pi)^d}\int_{\R^d}\widehat{\eta}(z,\mathbf{q})e^{i\mathbf{q}\cdot \rr}\,\dif \mathbf{q}.
\label{eq:app-eta-fourier}
\end{equation}

Recalling that $\eta=\bar{\eta}+\delta\eta$ yields
\begin{equation}
\partial_z\Wpsi+\pp\cdot\nabla_{\rr}\Wpsi
=
\Theta_{\bar{\eta}}[\Wpsi]
+
\Theta_{\delta\eta}[\Wpsi].
\label{eq:app-exact-split}
\end{equation}
Taking expectations finally leads to the desired equation
\begin{equation}
\partial_z\W(z,\rr,\pp)
+
\pp\cdot\nabla_{\rr}\W(z,\rr,\pp)
=
\Theta_{\bar{\eta}}[\W](z,\rr,\pp)
+
\mathcal{C}(z,\rr,\pp),
\label{eq:app-averaged}
\end{equation}
with the exact closure defect
\begin{equation}
\mathcal{C}(z,\rr,\pp)
:=
\E\!\left[\Theta_{\delta\eta}[\Wpsi](z,\rr,\pp)\right].
\label{eq:app-closure-defect}
\end{equation}

\section{Furutsu--Novikov closure in the white-noise limit}
\label{app:equivalence-FN}

This appendix explains how the exact closure defect introduced in
\eqref{eq:closure-defect-fourier} reduces, under a standard white-noise and Markov
approximation, to the nonlocal kinetic form used in the main text. The purpose is not to
derive a new stochastic model, but to show that the kinetic closure is consistent with the
usual Markovian moment equations obtained for random paraxial propagation.

For compactness, we introduce the following finite-difference operator
\begin{equation}
\mathcal{B}_{\mathbf{q}}[U](z,\rr,\pp)
=
e^{i\mathbf{q}\cdot\rr}
\left[
U\left(z,\rr,\pp-\frac{\mathbf{q}}{2k_0}\right)
-
U\left(z,\rr,\pp+\frac{\mathbf{q}}{2k_0}\right)
\right].
\label{eq:app-Bq-def}
\end{equation}
Then the random Wigner equation may be written in the form
\begin{equation}
\partial_z \Wpsi+\pp\cdot\nabla_{\rr}\Wpsi
=
\Theta_{\bar{\eta}}[\Wpsi]
+
\frac{i k_0}{(2\pi)^d}
\int_{\R^d}
\widehat{\delta\eta}(z,\mathbf{q})
\mathcal{B}_{\mathbf{q}}[\Wpsi](z,\rr,\pp)
\,\dif\mathbf{q}.
\label{eq:app-random-wigner-Bq}
\end{equation}
Accordingly,
\begin{equation}
\mathcal{C}(z,\rr,\pp)
=
\frac{i k_0}{(2\pi)^d}
\int_{\R^d}
\E\left[
\widehat{\delta\eta}(z,\mathbf{q})
\mathcal{B}_{\mathbf{q}}[\Wpsi](z,\rr,\pp)
\right]
\,\dif\mathbf{q}.
\label{eq:app-closure-Bq}
\end{equation}

We now assume that the unresolved refractive-index fluctuation is a centered Gaussian field,
white in the propagation variable and homogeneous in the transverse variables. In Fourier
variables this is written as
\begin{equation}
\E\left[
\widehat{\delta\eta}(z,\mathbf{q})
\widehat{\delta\eta}(z',\mathbf{q}')
\right]
=
(2\pi)^d
\delta(z-z')
\delta(\mathbf{q}+\mathbf{q}')
\widehat{R}_{\eta}(\mathbf{q}),
\label{eq:app-white-covariance}
\end{equation}
where $\widehat{R}_{\eta}(\mathbf{q})\ge 0$ is the transverse power spectral density of the
fluctuation. For any sufficiently regular functional $F[\delta\eta]$, the Furutsu--Novikov
formula gives
\begin{align}
\E\left[
\widehat{\delta\eta}(z,\mathbf{q})F[\delta\eta]
\right]
&=
\int
\E\left[
\widehat{\delta\eta}(z,\mathbf{q})
\widehat{\delta\eta}(z',\mathbf{q}')
\right]
\E\left[
\frac{\delta F}{\delta\widehat{\delta\eta}(z',\mathbf{q}')}
\right]
\,\dif z'\,\dif\mathbf{q}'.
\label{eq:app-FN-formula}
\end{align}
Applying \eqref{eq:app-FN-formula} to
$F=\mathcal{B}_{\mathbf{q}}[\Wpsi](z,\rr,\pp)$ yields
\begin{align}
\E\left[
\widehat{\delta\eta}(z,\mathbf{q})
\mathcal{B}_{\mathbf{q}}[\Wpsi](z,\rr,\pp)
\right]
&=
(2\pi)^d
\widehat{R}_{\eta}(\mathbf{q})
\E\left[
\frac{\delta\,\mathcal{B}_{\mathbf{q}}[\Wpsi](z,\rr,\pp)}
{\delta\widehat{\delta\eta}(z,-\mathbf{q})}
\right].
\label{eq:app-FN-applied}
\end{align}

It remains to approximate the instantaneous functional derivative. Differentiating
\eqref{eq:app-random-wigner-Bq} with respect to
$\widehat{\delta\eta}(z',\mathbf{q}')$ gives the following response equation
\begin{equation*}
\partial_z \mathcal{G}+\pp\cdot\nabla_{\rr}\mathcal{G}
=
\Theta_{\bar{\eta}}[\mathcal{G}] + \Theta_{\delta \eta}[\mathcal{G}]
+
\frac{i k_0}{(2\pi)^d} \delta(z-z')
\mathcal{B}_{\mathbf{q'}}[\Wpsi](z,\rr,\pp),
\end{equation*}
where $\mathcal{G}=\frac{\delta \Wpsi}{\delta \widehat{\delta \eta}}$.
In the Markov limit, only the
singular forcing at $z=z'$ contributes to the equal-time response. Thus, up to the conventional
equal-time factor associated with the chosen white-noise convention, one obtains
\begin{equation}
\frac{\delta\Wpsi(z,\rr,\pp)}
{\delta\widehat{\delta\eta}(z',\mathbf{q}')}
\simeq
\chi\,
\frac{i k_0}{(2\pi)^d}
\mathcal{B}_{\mathbf{q}'}[\Wpsi](z,\rr,\pp),
\label{eq:app-instant-response}
\end{equation}
where $\chi=1$ or $\chi=1/2$ depending on the convention used for the equal-time
white-noise response. This numerical factor may equivalently be absorbed into the definition of
the effective spectral density or of the scattering kernel. The important point is the operator
structure of the response.

Inserting \eqref{eq:app-instant-response} in \eqref{eq:app-FN-applied} leads to
\begin{equation}
\E\left[
\widehat{\delta\eta}(z,\mathbf{q})
\mathcal{B}_{\mathbf{q}}[\Wpsi](z,\rr,\pp)
\right]
\simeq
(2\pi)^d
\widehat{R}_{\eta}(\mathbf{q})
\chi\,
\frac{i k_0}{(2\pi)^d}
\E\left[
\mathcal{B}_{\mathbf{q}}\mathcal{B}_{-\mathbf{q}}[\Wpsi](z,\rr,\pp)
\right].
\label{eq:app-FN-BqBminusq}
\end{equation}
A direct computation shows that the phase factors cancel and
\begin{align}
\mathcal{B}_{\mathbf{q}}\mathcal{B}_{-\mathbf{q}}[U](\rr,\pp)
&=
2U(\rr,\pp)
-
U\left(\rr,\pp-\frac{\mathbf{q}}{k_0}\right)
-
U\left(\rr,\pp+\frac{\mathbf{q}}{k_0}\right).
\label{eq:app-Bq-composition}
\end{align}
Combining \eqref{eq:app-closure-Bq}, \eqref{eq:app-FN-BqBminusq}, and
\eqref{eq:app-Bq-composition}, the closure defect becomes
\begin{align}
\mathcal{C}(z,\rr,\pp)
&\simeq
\chi\,\frac{k_0^2}{(2\pi)^d}
\int_{\R^d}
\widehat{R}_{\eta}(\mathbf{q})
\left[
\W\left(z,\rr,\pp-\frac{\mathbf{q}}{k_0}\right)
+
\W\left(z,\rr,\pp+\frac{\mathbf{q}}{k_0}\right)
-
2\W(z,\rr,\pp)
\right]
\dif\mathbf{q}.
\label{eq:app-FN-closure-symmetric}
\end{align}
This is a closed nonlocal operator in momentum. Since
$\widehat{R}_{\eta}(\mathbf{q})=\widehat{R}_{\eta}(-\mathbf{q})$, it may also be written as a
gain--loss operator with momentum increment $\mathbf{s}=\mathbf{q}/k_0$:
\begin{equation}
\mathcal{C}(z,\rr,\pp)
\simeq
\int_{\R^d}
K_{\mathrm{FN}}(\mathbf{s})
\left[
\W(z,\rr,\pp+\mathbf{s})-\W(z,\rr,\pp)
\right]
\dif\mathbf{s},
\label{eq:app-FN-kinetic-form}
\end{equation}
where
\begin{equation}
K_{\mathrm{FN}}(\mathbf{s})
=
2\chi\,\frac{k_0^{d+2}}{(2\pi)^d}
\widehat{R}_{\eta}(k_0\mathbf{s}).
\label{eq:app-FN-kernel}
\end{equation}
The factor $2\chi$ depends only on the equal-time convention and is immaterial for the
hierarchical construction, since it is absorbed into the effective scattering kernel. Equation
\eqref{eq:app-FN-kinetic-form} is precisely the homogeneous nonlocal kinetic closure used in
the main text.

The small-angle Fokker--Planck limit follows by expanding
$\W(z,\rr,\pp+\mathbf{s})$ for kernels concentrated near $\mathbf{s}=0$. If the first moment of
$K_{\mathrm{FN}}$ vanishes by symmetry and the second moment is finite, then
\begin{equation}
\mathcal{C}(z,\rr,\pp)
\simeq
\frac12
\sum_{i,j=1}^d
A_{ij}\,\partial_{p_i}\partial_{p_j}\W(z,\rr,\pp),
\qquad
A_{ij}
=
\int_{\R^d}
s_i s_j K_{\mathrm{FN}}(\mathbf{s})\,\dif\mathbf{s}.
\label{eq:app-FN-diffusive-limit}
\end{equation}
In the isotropic case,
\begin{equation}
A_{ij}=2\Deff\,\delta_{ij},
\qquad
\Deff
=
\frac{1}{2d}
\int_{\R^d}
|\mathbf{s}|^2 K_{\mathrm{FN}}(\mathbf{s})\,\dif\mathbf{s}.
\label{eq:app-FN-Deff}
\end{equation}
Thus the white-noise Furutsu--Novikov closure leads to the same chain of reduced models as the one used in the main text.

This derivation also clarifies the status of the kinetic model. It is not introduced as an arbitrary phenomenological diffusion law. Under Gaussian, longitudinally white, transversely homogeneous fluctuations, it coincides with the Markovian second-order moment closure ~\cite{fannjiang2004scaling,fabbro2013scintillation,garnier2015white}. Away from this limit, the formula should be interpreted as an effective closure: the exact defect \eqref{eq:closure-defect-fourier} is replaced by a conservative nonlocal redistribution operator whose kernel sums up the unresolved medium--field correlations.

\section{Detailed Kramers--Moyal expansion and diffusive reduction}
\label{app:kramers-moyal}

We detail here the local diffusive reduction of the nonlocal kinetic closure
\begin{equation}
\partial_z \W(z,\rr,\pp)
+
\pp\cdot \nabla_\rr \W(z,\rr,\pp)
+
F(z,\rr,\pp)\cdot \nabla_\pp \W(z,\rr,\pp)
=
\Qkin[\W](z,\rr,\pp),
\label{eq:appB-kinetic}
\end{equation}
with jump operator
\begin{equation}
\Qkin[\W](z,\rr,\pp)
=
\int_{\R^d}
K(z,\rr;\mathbf{q},\pp)
\left[
\W(z,\rr,\pp+\mathbf{q})-\W(z,\rr,\pp)
\right]
\,\dif \mathbf{q}.
\label{eq:appB-jump}
\end{equation}
We assume that, for each fixed $(z,\rr,\pp)$, the kernel has finite first and second moments,
\begin{equation}
\int_{\R^d}|\mathbf{q}|\,K(z,\rr;\mathbf{q},\pp)\,\dif \mathbf{q}<\infty,
\qquad
\int_{\R^d}|\mathbf{q}|^2\,K(z,\rr;\mathbf{q},\pp)\,\dif \mathbf{q}<\infty,
\label{eq:appB-moment-assumptions}
\end{equation}
and, for a remainder estimate, also a finite third moment,
\begin{equation}
\int_{\R^d}|\mathbf{q}|^3\,K(z,\rr;\mathbf{q},\pp)\,\dif \mathbf{q}<\infty.
\label{eq:appB-third-moment}
\end{equation}
We further assume that $\W$ is sufficiently smooth in the momentum variable.

Expanding at second order $\W(z,\rr,\pp+\mathbf{q})$ around $\pp$ gives
\begin{equation}
\W(z,\rr,\pp+\mathbf{q})
=
\W(z,\rr,\pp)
+
\sum_{i=1}^d q_i\,\partial_{p_i}\W(z,\rr,\pp)
+
\frac12
\sum_{i,j=1}^d
q_iq_j\,\partial_{p_i}\partial_{p_j}\W(z,\rr,\pp)
+
\mathcal{R}_3(z,r,p,q),
\label{eq:appB-taylor}
\end{equation}
with integral remainder
\begin{equation}
\mathcal{R}_3(z,\rr,\mathbf{q},\pp)
=
\frac12
\sum_{i,j,k=1}^d
q_iq_jq_k
\int_0^1
(1-\theta)^2
\partial_{p_i}\partial_{p_j}\partial_{p_k}
\W(z,\rr,\pp+\theta \mathbf{q})\,\dif\theta.
\label{eq:appB-remainder}
\end{equation}
Substituting into \eqref{eq:appB-jump}, the zeroth-order terms cancel and one obtains
\begin{equation}
\Qkin[\W](z,\rr,\pp)
=
b(z,\rr,\pp)\cdot\nabla_\pp\W(z,\rr,\pp)
+
\frac12
\sum_{i,j=1}^d
A_{ij}(z,\rr,\pp)\,\partial_{p_i}\partial_{p_j}\W(z,\rr,\pp)
+
\mathcal{R}_{\mathrm{KM}}[\W](z,\rr,\pp),
\label{eq:appB-second-order}
\end{equation}
where
\begin{equation}
b_i(z,\rr,\pp)
:=
\int_{\R^d}q_i\,K(z,\rr;\mathbf{q},\pp)\,\dif \mathbf{q},
\label{eq:appB-bi}
\end{equation}
\begin{equation}
A_{ij}(z,\rr,\pp)
:=
\int_{\R^d}q_iq_j\,K(z,\rr;\mathbf{q},\pp)\,\dif \mathbf{q},
\label{eq:appB-Aij}
\end{equation}
and
\begin{equation}
\mathcal{R}_{\mathrm{KM}}[\W](z,\rr,\pp)
=
\int_{\R^d}
K(z,\rr;\mathbf{q},\pp)\,\mathcal{R}_3(z,\rr,\mathbf{q},\pp)\,\dif \mathbf{q}.
\label{eq:appB-km-remainder}
\end{equation}
Neglecting the remainder gives the second-order Kramers--Moyal truncation
\begin{equation}
\Qkin[\W](z,\rr,\pp)
\approx
b(z,\rr,\pp)\cdot\nabla_\pp\W(z,\rr,\pp)
+
\frac12
\sum_{i,j=1}^d
A_{ij}(z,\rr,\pp)\,\partial_{p_i}\partial_{p_j}\W(z,\rr,\pp).
\label{eq:appB-second-order-trunc}
\end{equation}

If the third derivatives of $\W$ are bounded in the relevant momentum region, then \eqref{eq:appB-remainder} yields the estimate
\begin{equation}
|\mathcal{R}_{\mathrm{KM}}[\W](z,\rr,\pp)|
\le
C_{\W}
\int_{\R^d}|\mathbf{q}|^3K(z,\rr;\mathbf{q},\pp)\,\dif \mathbf{q},
\label{eq:appB-remainder-bound}
\end{equation}
with $C_{\W}$ a constant that only depends on $\W$. This makes explicit the small-jump character of the approximation.

Substituting \eqref{eq:appB-second-order-trunc} into \eqref{eq:appB-kinetic} gives the local second-order closure
\begin{equation}
\partial_z \W
+
\pp\cdot \nabla_\rr \W
+
F(z,\rr,\pp)\cdot \nabla_\pp \W
=
b(z,\rr,\pp)\cdot \nabla_\pp \W
+
\frac12
\sum_{i,j=1}^d
A_{ij}(z,\rr,\pp)\,\partial_{p_i}\partial_{p_j}\W.
\label{eq:appB-local-nondiv}
\end{equation}
Equivalently, the first-order term may be moved to the left-hand side and interpreted as a correction to the deterministic momentum drift.

In the symmetric case
\begin{equation}
K(z,\rr;\mathbf{q},\pp)=K(z,\rr;-\mathbf{q},\pp),
\label{eq:appB-symmetry}
\end{equation}
the first jump moment vanishes,
\begin{equation}
b(z,\rr,\pp)=0,
\label{eq:appB-b-zero}
\end{equation}
so that
\begin{equation}
\Qkin[\W]
\approx
\frac12
\sum_{i,j=1}^d
A_{ij}(z,\rr,\pp)\,\partial_{p_i}\partial_{p_j}\W.
\label{eq:appB-symmetric-second-order}
\end{equation}
If, in addition, the kernel is isotropic in the jump variable, then
\begin{equation}
A_{ij}(z,\rr,\pp)=2D(z,\rr,\pp)\,\delta_{ij},
\label{eq:appB-isotropic-A}
\end{equation}
with
\begin{equation}
D(z,\rr,\pp)
=
\frac{1}{2d}
\int_{\R^d}|\mathbf{q}|^2K(z,\rr;\mathbf{q},\pp)\,\dif \mathbf{q}.
\label{eq:appB-D-general}
\end{equation}
Hence the local closure reduces to
\begin{equation}
\Qkin[\W]\approx D(z,\rr,\pp)\,\Delta_\pp\W.
\label{eq:appB-isotropic-diff}
\end{equation}

Finally, if the effective medium is homogeneous at the closure level, one may take $D(z,\rr,\pp)\equiv \Deff$ constant, and the reduced model becomes
\begin{equation}
\partial_z \W(z,\rr,\pp)
+
\pp\cdot \nabla_\rr \W(z,\rr,\pp)
=
\Deff\,\Delta_\pp \W(z,\rr,\pp),
\label{eq:appB-minimal-fp}
\end{equation}
with
\begin{equation}
\Deff
=
\frac{1}{2d}
\int_{\R^d}|\mathbf{q}|^2K(\mathbf{q})\,\dif \mathbf{q}.
\label{eq:appB-Deff}
\end{equation}
In dimension $d=2$,
\begin{equation}
\Deff
=
\frac14
\int_{\R^2}|\mathbf{q}|^2K(\mathbf{q})\,\dif \mathbf{q},
\label{eq:appB-Deff-d2}
\end{equation}
and for a radial kernel $K(\mathbf{q})=K(|\mathbf{q}|)$,
\begin{equation}
\Deff
=
\frac{\pi}{2}
\int_0^\infty \rho^3K(\rho)\,\dif \rho.
\label{eq:appB-Deff-radial}
\end{equation}

Thus, the local Fokker--Planck closure is the second-order small-jump approximation of the nonlocal kinetic operator: the drift term is controlled by the first jump moment, the diffusion tensor by the second jump moment, and the minimal homogeneous isotropic model follows when the kernel is symmetric, isotropic, and effectively constant in $(z,\rr,\pp)$.


\bibliographystyle{elsarticle-num}
\bibliography{references}

\end{document}